\documentclass[journal,letterpaper,twocolumn,final,10pt]{IEEEtran}

\usepackage{amsmath}
\usepackage{amsfonts}
\usepackage{breqn}
\usepackage{dsfont}
\usepackage{mathrsfs}
\usepackage{pifont}
\usepackage{amssymb}
\usepackage{verbatim}
\usepackage{upgreek}
\usepackage{color}
\usepackage{graphicx}
\usepackage{algorithmic}
\usepackage{algorithm}
\usepackage{epsfig}
\usepackage{eucal}
\usepackage{cite}
\usepackage{subfigure}

\DeclareMathOperator*{\argmax}{arg\,max}

\newcommand{\bbm}{\begin{bmatrix}}
\newcommand{\ebm}{\end{bmatrix}}

\newcommand{\bit}{\begin{itemize}}
\newcommand{\eit}{\end{itemize}}

\newcommand{\ben}{\begin{enumerate}}
\newcommand{\een}{\end{enumerate}}

\newcommand{\bdesc}{\begin{description}}
\newcommand{\edesc}{\end{description}}

\newcommand{\bea}{\begin{array}}
\newcommand{\eea}{\end{array}}

\newcommand{\beqa}{\begin{eqnarray}}
\newcommand{\eeqa}{\end{eqnarray}}

\newcommand{\ds}{\displaystyle}

\newcommand{\Comment}[1]{}

\def\R{{\mathds R}}
\def\Real{{\mathfrak Re}}
\def\Imag{{\mathfrak Im}}
\def\C{{\mathds C}}

\def\cC{\mbox{$\CMcal C$}}

\def\cG{\mbox{$\mathcal G$}}

\def\cN{\mbox{$\CMcal N$}}

\newcommand{\be}{\begin{equation}}
\newcommand{\ee}{\end{equation}}

\newcommand{\bzero}{{\mbox{\boldmath $0$}}}

\newcommand{\bc}{{\mbox{\boldmath $c$}}}

\newcommand{\bm}{{\mbox{\boldmath $m$}}}

\newcommand{\bt}{{\mbox{\boldmath $t$}}}

\newcommand{\bx}{{\mbox{\boldmath $x$}}}
\newcommand{\by}{{\mbox{\boldmath $y$}}}

\newcommand{\bz}{{\mbox{\boldmath $z$}}}

\newcommand{\bG}{{\mbox{\boldmath $G$}}}
\newcommand{\bH}{{\mbox{\boldmath $H$}}}
\newcommand{\bI}{{\mbox{\boldmath $I$}}}

\newcommand{\bM}{{\mbox{\boldmath $M$}}}

\newcommand{\bT}{{\mbox{\boldmath $T$}}}

\newcommand{\bX}{{\mbox{\boldmath $X$}}}

\newcommand{\bZ}{{\mbox{\boldmath $Z$}}}

\newcommand{\diag}{\mbox{\boldmath\bf diag}\, }

\newcommand{\Var}{{\bf\sf Var}}

\newcommand{\bSigma}{{\mbox{\boldmath $\Sigma$}}}

\newcommand{\bsigma}{{\mbox{\boldmath $\sigma$}}}

\newcommand{\test}{\mbox{$
\begin{array}{c}
\stackrel{ \stackrel{\textstyle H_1}{\textstyle >} }{
\stackrel{\textstyle <}{\textstyle H_0} }
\end{array}
$}}

\title{Radar Adaptive Detection Architectures for Heterogeneous Environments}

\author{Jun Liu, \IEEEmembership{Senior Member, IEEE}, Davide Massaro,  Danilo Orlando, \IEEEmembership{Senior Member, IEEE}, 
Alfonso Farina, \IEEEmembership{Life Fellow, IEEE}
\thanks{Jun Liu is with the Department of Electronic Engineering and Information Science, University of Science and Technology of China, Hefei 230027, 
China. E-mail: {\tt junliu@ustc.edu.cn}.}
\thanks{Davide Massaro is with Elettronica S.p.A., Via Tiburtina Valeria km 13,700, 00131 Roma, Italy. E-mail: {\tt davide.massaro3@gmail.com}.}
\thanks{D. Orlando is with the Engineering Faculty of Universit\`a degli Studi ``Niccol\`o Cusano'', via Don Carlo Gnocchi 3, 00166 Roma, Italy. E-mail: {\tt danilo.orlando@unicusano.it}.}
\thanks{Alfonso Farina is a Technical Consultant (previously with Selex ES), Via Helsinki 14, Rome, Italy. E-mail: {\tt alfonso.farina@outlook.it}.}
}

\begin{document}

\maketitle

\begin{abstract}
In this paper, four adaptive radar architectures for target detection in heterogeneous Gaussian environments 
are devised. The first architecture relies on a cyclic optimization exploiting the Maximum Likelihood 
Approach in the original data domain, whereas the second detector is a function of transformed data 
which are normalized with respect to their energy and with the unknown parameters estimated through 
an Expectation-Maximization-based alternate procedure. 
The remaining two architectures are obtained by suitably combining the estimation procedures and 
the detector structures previously devised. Performance analysis, conducted on both simulated and 
measured data, highlights that the architecture working in the transformed domain guarantees the constant false 
alarm rate property with respect to the interference power variations and a limited detection loss with respect
to the other detectors, whose detection thresholds nevertheless are very sensitive to the interference power.
\end{abstract}


\section{Introduction}\label{Sec:Introduction}
Last-generation radar systems are provided with a considerable abundance of 
computation power, which was inconceivable a few decades ago.
As a consequence, more and more sophisticated processing schemes are being incorporated 
into radar systems as corroborated by the novel architectures which continuously 
appear in the open literature. Such architectures provide enhanced performances at the price of an increased computational load \cite{kelly1986adaptive,robey1992cfar,BOR-Morgan,JunLiu00,CP00,HaoSP_HE}. 
A common issue concerning the design of these architectures is related to the statistical assumptions for 
the interference affecting the set of data to be processed, which consists of the 
range Cell Under Test (CUT) and an additional cluster of data, obtained collecting echoes in proximity 
of the CUT and used for estimation purposes. 
For instance, in the case of Gaussian interference,
the additional cluster of data, also known as secondary data set, is assumed to share
the same spectral properties of the interference as that in the CUT. This situation is referred to as homogeneous environment,
which is widespread in the radar community \cite{Richards,kelly1986adaptive,robey1992cfar,BOR-Morgan,DD,
Ciuonzo2,Raghavan} and represents the ``entry-level'' interference model in the design of adaptive decision schemes. 
Under the homogeneous environment, secondary data are exploited to obtain reliable estimates of either the 
interference covariance matrix (raw space-time data processing) or the interference power (after space and/or time beamforming) \cite{reed1974rapid}. 
Then, such estimates
are plugged into decision statistics to achieve adaptivity and, more importantly, the Constant False Alarm Rate (CFAR) property. 
It is relevant to underline that the detection performance strongly depends on the estimation quality of the unknown
interference parameters, which, in turn, is tied to the amount of secondary data (or, more precisely,
to the available information carried by them). However, the presence of inhomogeneities in the secondary 
data generates a severe performance degradation for those architectures designed under 
the homogeneous environment \cite{Melvin-2000} and the CFAR property is no longer ensured. 
Indeed, secondary data are often contaminated by power variations over range, clutter discretes, 
and other outliers, which drastically reduce the number of homogeneous secondary data. Furthermore, in target-rich environments
structured echoes in secondary data can overnull the signal of interest and result in missed detections \cite{bergin2002gmti}.

In the open literature, there exist a plethora of approaches to cope with small volumes of homogeneous 
training samples. For instance, the knowledge-aided paradigm represents an effective means to obtain 
reliable estimates in sample-starved scenarios. It consists in accounting for the available {\em a priori}
information at the design stage \cite{HaoSP_HE,Hongbin,DeMaioFoglia01}. Alternatively, 
{\em ad hoc} decision rules can be designed by forcing the same properties as the Generalized Likelihood Ratio 
Test \cite{YuriJohnson} or using the expected-likelihood \cite{yuriBesson}. 
Other widely used techniques consist in the regularization (or shrinkage) 
of the sample covariance matrix towards a given matrix \cite{WieselHero,Tyler,yuriModified,gerlach1} or
in detecting and suppressing the outliers in order to make the training set homogeneous
\cite{629144,1191096,767347,7575566,851934,1433140}.
Finally, the homogeneous model can be suitably extended to account for heterogeneous data.
Among the frequently used assumptions to depict a non-homogeneous
scenario there is the Partially Homogeneous Environment (PHE), where 
both the CUT and secondary data share the same interference covariance
matrix structure but different interference power levels \cite{kraut1999cfar}. Though keeping a
relative mathematical tractability, the PHE leads to an increased robustness to inhomogeneities
since the assumed difference in power level accounts for terrain type variations, height profile, and shadowing which may 
appear in practice \cite{Ward1994}. Additionally, the PHE subsumes the homogeneous environment as a special case.

In this paper, we address the problem of detecting point-like targets in heterogeneous scenarios by extending the 
PHE to account for interference power variations between consecutive samples. Specifically, for each range bin, 
the system collects the echoes due to the transmission of a coherent burst of pulses. Such echoes are characterized 
by different interference power levels (nonstationary random process) leading to a ``Fully-Heterogeneous'' Gaussian Environment (denoted in the following by the acronym HE). 
Under these assumptions, we design four adaptive architectures which do not use secondary data and that
represent different ways of solving the same detection problem.

The first architecture is devised in the original data domain
exploiting the Likelihood Ratio Test (LRT) where the unknown target and interference parameters 
are estimated resorting to a cyclic optimization based upon the Maximum Likelihood Approach (MLA).
This alternating estimation approach is dictated by the fact that
the straightforward application of the MLA is a difficult task for the estimation problem at hand.
Moreover, it is important to underline that in this case the CFAR property cannot be a priori predicted 
and an analysis is required to ascertain the sensitivity of the detection threshold to the interference 
power variations. On the other hand, the second proposed architecture relies on transformed data. 
The line of reasoning behind this transformation resides in the fact that the joint probability density 
function (pdf) of the modulus and phase of a complex normal random variable (rv) with zero mean and 
variance $\sigma^2>0$ (i.e., the data distribution under the null hypothesis) is given by the product 
between the pdf of a Rayleigh rv with parameter $\sigma^2/2$
by that of a rv uniformly distributed between $0$ and $2\pi$ \cite{papoulis2002probability} that, clearly, 
does not depend on $\sigma^2$. As a consequence, normalizing the considered complex 
normal random variable with respect to its modulus leads to a distribution independent of $\sigma^2$.
With this remark in mind, the original data can be transformed in order to get rid of the dependence on
the variance at least under $H_0$ paving the way to the design of CFAR decision rules.
Remarkably, this idea can be framed in a more general context by invoking the {\em Invariance Principle} \cite{LehmannBook} and the so-called {\em Directional Statistics} \cite{MardiaJupp} in order to also account for 
normalized random variables with nonzero mean.

To be more definite, the Invariance Principle allows us to prove that
data normalized with respect to their energy represent
a Maximal Invariant Statistic (MIS) which is functionally independent of 
scaling factors (namely, of the interference power levels) under the noise-only hypothesis.
As a consequence, any decision rule based upon the MIS is invariant to interference power variations
ensuring the CFAR property with respect to the latter.
In addition, the distribution of the normalized data under the target-plus-noise hypothesis is obtained by 
exploiting the directional statistics and, 
in the specific case, the {\em Angular Gaussian} distribution. In this framework,
we devise a decision scheme based upon the LRT, which represents the main 
technical novelty of this paper (at least to the best of the authors' knowledge). 
Specifically, note that in this case a cyclic estimation procedure based upon 
MLA (as in the previous case) cannot be applied as under the alternative hypothesis the pdf 
of the normalized data has an expression that is very difficult to handle from 
a mathematical point of view. For this reason, we still use an alternating optimization procedure but
we replace the MLA with the Expectation Maximization (EM) algorithm \cite{Dempster77} specialized for the 
exponential family, since it is a simple iterative algorithm that provides
closed-form updates for the parameter estimates at each step and reaches at least a local stationary point.
However, the application of the EM algorithm requires the presence of hidden data. To this end,
we disregard that original data (referred to in the EM framework as {\em complete data}) are available and fictitiously
assume that only normalized data can be processed whereas data norms are the hidden variables.
Remarkably, we expect that the architecture developed under the above framework, by virtue of 
the Invariance Principle, guarantees the CFAR property with respect to the interference power level. 
Finally, the third and fourth decision schemes, referred to as {\em cross} architectures, are obtained
by combining the estimates provided by the MLA-based cyclic procedure with the LRT of the transformed data
and the estimates provided by the EM-based alternating procedure with the LRT of the original data, respectively.
It is clear that also for these architectures a CFAR analysis is required to ascertain 
their sensitivity to the interference power variations.

The numerical examples are built up resorting to simulated and real recorded data.
More precisely, the nominal behavior of the proposed architectures is investigated over 
simulated data which adhere to the design assumptions. This analysis
confirms the expected behavior in terms of CFARness of the second architecture. 
On the other hand, the remaining detectors using original data
are very sensitive to the interference power variations, even though two of them ensure better
detection performance than the invariant detector. Finally, the results observed for 
simulated data are corroborated by testing the proposed architectures on data
collected in winter 1998 using the McMaster IPIX radar in Grimsby, on the shore of Lake Ontario, between Toronto
and Niagara Falls \cite{haykin2002uncovering}.

The remainder of this paper is organized as follows. In the next section, we formulate the detection 
problem in both the original data domain and invariant domain. In Section \ref{Sec:DetectorDesign}, we 
describe the procedures to estimate the unknown parameters and devise the LRT-based adaptive architectures, 
while Section \ref{Sec:IllustrativeExamples} contains illustrative examples. Finally, in 
Section \ref{Sec:Conclusions}, we draw the conclusions and point out future research tracks. Some mathematical derivations are confined to the appendices.

\subsection{Notation}
In the sequel, vectors and matrices are denoted by boldface lower-case and upper-case letters, respectively. Symbol $(\cdot)^T$ denotes transpose. 
For a generic vector $\bx$, symbol $\|\bx\|$ indicates its Euclidean norm.
$\R$ is the set of real numbers, $\R^{N\times M}$ is the Euclidean space of $(N\times M)$-dimensional real 
matrices (or vectors if $M=1$), $\R_+^{N\times M}$ is the set of $(N\times M)$-dimensional 
real matrices (or vectors if $M=1$) whose entries are greater than or equal to zero, and $\C$ is the set of complex numbers.
If $\bx$ is a generic $N$-dimensional vector then $\diag(\bx)$ is $N\times N$-dimensional diagonal matrix whose nonzero entries are the elements of $\bx$.
Symbols $\Gamma(\cdot)$ and $\odot$ denote the Eulerian Gamma function and the element-wise Hadamard product, respectively.
Symbols $\Real\left\{ z \right\}$ and $\Imag\left\{ z \right\}$ indicate the real and imaginary parts of the complex number $z$, respectively. 
$\bI_N$ stands for the $N \times N$ identity matrix, while $\bzero$ is the null vector or matrix of proper dimensions. 
Let $\bx$ and $\by$ be two random vectors, then $E\left[\bx |\by\right]$ and $\Var\left[\bx |\by\right]$ are the conditional expectation and the conditional variance of $\bx$ given $\by$, respectively. 
Finally, we write $\bx\sim\cC\cN_N(\bm, \bM)$ if $\bx$ is a complex circular $N$-dimensional normal vector with mean $\bm$ and positive definite 
covariance matrix $\bM$, $\bx\sim \cN_N(\bm, \bM)$ if $\bx$ is an $N$-dimensional normal vector with mean $\bm$ and positive definite 
covariance matrix $\bM$, $x\sim U(a,b)$ is $x$ is a random uniform variable ranging in the interval $[a,b]$.

\section{Problem Formulation}
\label{Sec:ProblemFormulation}
Let us consider a radar system that transmits a coherent burst of $K$ pulses to sense the surrounding environment. 
The backscattered signal impinging the radar undergoes a baseband down-conversion and a filtering matched 
to the transmitted pulse waveform. Then, the output of the matched filter is suitably sampled in order to 
form the range bins.
In the case where the system is equipped with $N$ spatial channels, the samples from each channel are combined using suitable 
weights in a digital beamformer \cite{Richards,antennaBased}.
Summarizing, for each range bin, $K$ complex samples are available (slow-time), which result 
from the superposition between an interference component and
a possible useful signal component. When the former is stationary over the range and/or time dimension, 
a set of training 
samples (secondary data) in proximity to that under test can be exploited
to come up with adaptive decision schemes capable of ensuring 
the CFAR property \cite{Richards,kelly1986adaptive,robey1992cfar,BOR-Morgan}. 
However, in practice there exist situations where the conventional approach based upon the secondary 
data set might fail due to the presence of interference power variations over range (fast-time) and 
pulses (slow-time), clutter discretes, and other outliers.
As a consequence, interference within secondary data is no longer representative of that in the CUT and
architectures designed for the homogeneous environment exhibit a significant performance degradation. 
More importantly, the CFAR property is no longer ensured \cite{Melvin-2000}.

To face with the above situations, in what follows, we focus on the HE and assume that, at the design level, 
interference affecting the $K$ samples exhibits 
different power levels. Specifically, let us denote by $x_1,\ldots,x_K\in\C$ the complex returns (at the output
of the beamformer) representative of the CUT and focus on the problem of deciding whether or 
not they contain useful signal components, 
which can be formulated in terms of the following hypothesis test
\be\label{eqn:original_Problem}
\left\{
\begin{array}{ll}
H_0: \ x_k\sim\cC\cN_1(0,2\sigma_k^2), & k=1,\ldots,K,
\\
H_1: \ x_k\sim\cC\cN_1(\alpha,2\sigma_k^2), & k=1,\ldots,K,
\end{array}
\right.
\ee
where\footnote{The factor $2$ is used to simplify the notation.} $\sigma_k^2\geq C_0>0$, $k=1,\ldots,K$, is the power 
of the interference affecting the echo associated with the $k$th transmitted pulse; 
$\alpha\in\C$ accounts for target response and channel 
effects\footnote{Note that the behavior of target and channel is assumed stationary in time.};
$x_k$s are assumed statistically independent.
As for $C_0$, it is a positive constant that accounts for the minimum allowable power level of the interference.
This lower bound has been introduced for {\em regularization} purposes. As a matter of fact,
note that the number of unknown parameters in \eqref{eqn:original_Problem} is $K+2$, 
namely $\sigma^2_k$, $k=1,\ldots,K$, $\Real\{\alpha\}$, and $\Imag\{\alpha\}$, whereas the number 
of available data is $2K$, i.e., $\Real\{x_k\}$ and $\Imag\{x_k\}$, $k=1\ldots,K$. Even 
though $2K>K+2$ when $K>2$, the problem of estimating $\sigma_k^2$ is ill-conditioned due to
the small amount of data sharing the same $\sigma_k^2$. Thus, a prospective estimator of $\sigma_k^2$
should exhibit a significant variance that can be limited by forcing the mentioned lower bound.
Finally, in practice $C_0$ could be set according to the level of the system internal noise, which can be
estimated by collecting noisy samples when the antenna is disengaged by means of a switch
(or circulator) device.

Problem \eqref{eqn:original_Problem} can be recast in terms of $2$-dimensional Gaussian vectors whose entries are the real and imaginary parts of the
complex samples, namely
\be
\bx_k= \begin{bmatrix} \Real\left\{ x_k \right\}, \Imag\left\{ x_k \right\} \end{bmatrix}^T \in \R^{2 \times 1}, \ k = 1,\cdots, K,
\ee
which, by definition, obey the $2$-variate Gaussian distribution with mean $\bzero$ and
$\bm=[\Real\left\{ \alpha \right\}, \ \Imag\left\{ \alpha \right\}]^T$ under $H_0$ and $H_1$, respectively. 
The covariance matrix is $\sigma_k^2\bI_2$ under both hypotheses (this is a straightforward consequence
of the definition of complex circular Gaussian random variable). 
It follows that \eqref{eqn:original_Problem} is equivalent to
\be\label{eqn:detectionProblem_vector}
\begin{cases}
H_0: \bx_k \sim \cN_2\left(\bzero, \ds{\sigma_k^2}\bI_2\right), & k = 1,\ldots,K,
\\
H_1: \bx_k \sim \cN_2\left(\bm, \ds{\sigma_k^2}\bI_2\right), & k = 1,\ldots,K,
\end{cases}
\ee
and the pdf of $\bx_k$ under $H_i$, $i=0,1$, is given by
\be
f_{x,i}(\bx_k; i\bm,\sigma^2_k) = \frac{1}{2\pi \sigma^2_k}\exp\left\{ -\frac{\|\bx_k-i\bm\|^2}{2\sigma^2_k} \right\}.
\ee
The design of CFAR decision rules for the above problem, where data experience interference power variations,
might represent a difficult task. For this reason, we transform data in order to remove the dependence of data distribution on $\sigma^2_k$s under $H_0$. 
In fact, as stated in Section \ref{Sec:Introduction}, normalizing a zero-mean complex normal
rv with respect to its modulus makes the resulting distribution independent of its variance. However, under
$H_1$, due to the nonzero mean, it is more suitable to frame the next developments in the context of the
{\em Directional Statistics} \cite{MardiaJupp}.
Such statistics can be obtained by transforming $\bx_k$, $k=1,\ldots,K$, into unit-norm vectors. 
As a consequence, any decision statistic, which is a function of the transformed data, naturally gets the 
CFAR property with respect to $\sigma^2_k$s. This behavior can be formally explained in the 
context of the Theory of Invariance \cite{LehmannBook}, 
which requires the identification of a suitable
group of transformations.
More precisely, let us define the set of vectors 
$\cC = \left\{ \bc\in\R_+^{K\times 1} \right\}$ along with 
the composition operator ``$\circ$'' defined as $\forall \bc_1,\bc_2\in\cC: \ \bc_1 \circ \bc_2 = \bc_1 \odot \bc_2$.
Then, it is not difficult to show that $\cG=(\cC, \circ)$ constitutes a group, since it satisfies the following elementary axioms
\begin{itemize}
\item $\cG$ is \emph{closed} with respect to the operation defined in the last equation;
\item $\forall \bc_1$, $\bc_2$, and
$\bc_3\in \cG$: $[ \bc_1 \circ \bc_2]\circ\bc_3=\bc_1\circ[\bc_2\circ\bc_3]$
(\emph{associative property});
\item there exists a unique $\bc_I\in\cG$ such
that $\forall \bc\in\cG$: $\bc_I\circ \bc=\bc\circ \bc_I=\bc$
(\emph{existence of the identity element});
\item $\forall  \bc \in \cG$, there exists $\bc_{-1}\in\cG$
such that $\bc_{-1}\circ\bc=\bc\circ\bc_{-1}=\bc_I$
(\emph{existence of the inverse element}).
\end{itemize}
Besides, it is evident that this group preserves the family of distributions and modifies 
the scaling factors under the action $\bG(\cdot,\ldots,\cdot)$ defined as 
$\bG(\bx_1,\ldots,\bx_K) = [\bc(1)\bx_1,\ldots,\bc(K)\bx_K]$.
Thus, exploiting the Principle of Invariance, we can replace the original data vectors with a suitable 
function of them, namely the MIS, which is functionally invariant to the considered group of transformations.
As a result, under $H_0$ the statistical dependence on $\sigma^2_i$ is removed. In Appendix \ref{maximal_inv_proof} it is shown that a MIS with respect to $\cG$ is given by
\be
\bT\left(\bx_1,\ldots,\bx_K\right) = \left[\bz_1,\ldots,\bz_K\right],
\label{eqn:maxinv}
\ee
where $\bz_k= \frac{\bx_k}{\|\bx_k\|} \in \R^{2 \times 1}$, $k=1,\ldots,K$,
which, evidently, only depend on the direction of $\bx_k$ in $\R^{2\times 1}$. 

Thus, in the invariant domain, the detection problem at hand can be written as
\be
\begin{cases}
H_0: \bz_k \sim f_0(\bz_k), & k = 1,\ldots,K,
\\
H_1: \bz_k \sim f_1(\bz_k;\bm,\sigma^2_k), & k = 1,\ldots,K,
\end{cases}
\label{eqn:detectionProblem_invariant}
\ee
where vectors $\bz_k$, $k=1,\ldots,K$, obey the {\em Angular Normal Distribution} \cite{MardiaJupp} with pdfs:
$f_0(\bz_k) = {1}/{(2\pi)}$ and (as shown in Appendix \ref{App:derivation_pdf_H1})
\be
f_1(\bz_k;\bm,\sigma^2_k) = 
\frac{\exp\left\{-\frac{\|{\scriptsize \bm}\|^2}{2\sigma^2_k}\right\}}{2\pi} \left[1+\frac{\ds\frac{\bz_k^T \bm}{\sigma_k} \Phi\left(\frac{\bz_k^T \bm}{\sigma_k}\right)}{\ds\varphi\left(\frac{\bz_k^T \bm}{\sigma_k}\right)}\right],
\label{eqn:pdf_H1}
\ee
under $H_0$ and $H_1$, respectively. In \eqref{eqn:pdf_H1}, $\Phi(\cdot)$ and $\varphi(\cdot)$ are the 
Cumulative Distribution Function (CDF) and the pdf of a standard Gaussian random variable, respectively. 
Finally, note that $f_1(\bz_k;\bzero,\sigma_k)=f_0(\bz_k)$ and, hence, the formal structure of the
detection problem at hand given by
\be
H_0: \ \bm=\bzero, \quad H_1: \ \bm\neq\bzero,
\label{eqn:formalStructure}
\ee
remains unaltered.

Detectors designed in this domain are expected to ensure the CFAR property as corroborated by the analysis
presented in Section IV.

\section{Detector Design}
\label{Sec:DetectorDesign}
In this section, we devise adaptive detection architectures for problem \eqref{eqn:formalStructure} 
exploiting data from either the original domain, or the invariant domain, or both domains. To this end, we 
resort to the LRT where the unknown parameters under each hypothesis are replaced by suitable
estimates. Specifically, the architectures operating in one domain are formed by coupling the LRT and parameter
estimates for the same domain, whereas those based upon data from both domains, namely the cross architectures, 
are built up by plugging the estimates obtained in one domain into the LRT for the other domain and vice versa.

As for the design methodology, it is important to observe that under the assumptions 
considered in Section \ref{Sec:ProblemFormulation} 
the plain MLA approach does not represent a viable route towards the estimation of the unknown parameters 
$\bm$ and $\bsigma^2$ as it requires to solve mathematically intractable equations in both domains 
(at least to the best of authors' knowledge). For this reason, we resort to a cyclic optimization paradigm \cite{Stoica_alternating}, which consists in partitioning the parameter set into two suitable subsets 
and, at each iteration, in estimating the parameters of a subset assuming the 
other parameters known. In the original domain, at each iteration of this procedure 
the application of the MLA is practicable, while in the transformed domain the MLA 
still leads to difficult equations. In order to cope with this
drawback, we resort to the EM approach \cite{Dempster77}, which, as already stated, is an iterative algorithm 
providing closed-form updates for the sought estimates. Now, the application of the EM algorithm requires 
the presence of hidden variables in addition to observed data. Therefore, we fictitiously assume that original 
data are no longer available and, hence, that data norms represent the hidden variables.

Finally, before proceeding with the decision rule designs, for future reference, 
let us define $\bX=[\bx_1,\ldots,\bx_K]$, 
$\bZ=[\bz_1,\ldots,\bz_K]$, and $\bsigma^2=[\sigma^2_1,\ldots,\sigma^2_K]^T$.

\subsection{Original Data Domain}
This subsection is devoted to the derivation of an adaptive architecture whose decision statistic is a function of $\bX$.
To this end, the unknown parameters under $H_1$ are estimated by means of a procedure combining the ML approach with a cyclic optimization method \cite{Stoica_alternating}. On the other hand, under $H_0$, we compute the ML estimate of $\bsigma^2$.

Let us begin with the expression of the LRT
\be
\Psi_1(\bX;\bm,\bsigma^2)=\frac{\ds f_{X,1}(\bX;\bm,\bsigma^2)}
{\ds f_{X,0}(\bX;\bzero,\bsigma^2)}\test\eta,
\label{eqn:LRT_original}
\ee
where $f_{X,1}(\bX;\bm,\bsigma^2)=\prod_{k=1}^K f_{x,1}(\bx_k;\bm,\sigma^2_k)$, 
$f_{X,0}(\bX;\bzero,\bsigma^2)=\prod_{k=1}^K f_{x,0}(\bx_k;\bzero,\sigma^2_k)$,
$\eta$ is the threshold\footnote{Hereafter, the generic detection threshold is denoted by $\eta$.} to be set 
in order to guarantee the required Probability of False Alarm ($P_{fa}$); parameters $\bm$ and $\bsigma^2$ 
have to be estimated from $\bX$ in order to make the above decision rule adaptive.

Under $H_0$, the unknown parameters are estimated as follows
\be
\widehat{\bsigma}^2_0=\argmax_{\sigma_k^2\geq C_0 \atop k=1,\ldots,K}\prod_{k=1}^K f_{x,0}(\bx_k;\bzero,\sigma^2_k).
\ee
Thus, setting to zero the first derivative of $\prod_{k=1}^K f_{x,0}(\bx_k;0,\sigma^2_k)$ with respect 
to $\sigma_k^2$ and accounting for the constraint $\sigma_k^2\geq C_0$, we obtain that
\be
\widehat{\bsigma}^2_0=
\begin{bmatrix}
\max\{\frac{1}{2}\|\bx_1\|^2,C_0\}
\\
\vdots
\\
\max\{\frac{1}{2}\|\bx_K\|^2,C_0\}
\end{bmatrix}
=\begin{bmatrix}
\widehat{\sigma}^2_{0,1}
\\
\vdots
\\
\widehat{\sigma}^2_{0,K}
\end{bmatrix}.
\ee
As for the estimation under $H_1$, we proceed according to the following rationale
\begin{enumerate}
\item assume that $\bsigma^2$ is known and compute the resulting ML estimate of $\bm$;
\item replace $\bm$ with the estimate obtained at the previous step and derive the ML 
estimate of $\bsigma^2$ with the constraint $\sigma_k^2\geq C_0$, $k=1,\dots,K$;
\item repeat the above steps until a stopping criterion is satisfied.
\end{enumerate}
As for the first step, it is not difficult to show that the ML estimate of $\bm$ when 
$\bsigma^2$ is equal to an initial value, $\bar{\bsigma}^2$ say, has the following expression
\be
\widehat{\bm}=\left[ \sum_{k=1}^K \frac{1}{\bar{\sigma}_k^2} \right]^{-1} 
\sum_{k=1}^K\frac{\bx_k}{\bar{\sigma}_k^2},
\label{eqn:m_alternateEstimate}
\ee
whereas the estimate of $\bsigma^2$ when $\bm=\widehat{\bm}$ (second step) is given by
\be
\widehat{\bsigma}^2_1=
\begin{bmatrix}
\max\{\frac{1}{2}\|\bx_1-\widehat{\bm}\|^2,C_0\}
\\
\vdots
\\
\max\{\frac{1}{2}\|\bx_K-\widehat{\bm}\|^2,C_0\}
\end{bmatrix}=
\begin{bmatrix}
\widehat{\sigma}^2_{1,1}
\\
\vdots
\\
\widehat{\sigma}^2_{1,K}
\end{bmatrix}.
\label{eqn:sigmaEstimateOriginal_reg}
\ee
It is important to observe that $C_0$ prevents \eqref{eqn:m_alternateEstimate} from diverging, 
since there could exist an index $\tilde{k}$ such that $\bx_{\tilde{k}}-\widehat{\bm}\approx \bzero$.

Finally, the estimate updates terminate when a stopping criterion is satisfied. 
Specifically, let us denote by $\widehat{\bm}^{(n)}$, $(\widehat{\bsigma}^2_1)^{(n)}$,
$\widehat{\bm}^{(n-1)}$, and $(\widehat{\bsigma}^2_1)^{(n-1)}$ the available estimates 
at the $n$th and $(n-1)$th iterations, respectively, then 
the alternating procedure terminates when
\be
\left| f_{X,1}(\bX;\widehat{\bm}^{(n)}\!\!,(\widehat{\bsigma}^2_1)^{(n)}) \!
- \! f_{X,1}(\bX;\widehat{\bm}^{(n-1)}\!\!,(\widehat{\bsigma}^2_1)^{(n-1)}) 
\right|<\epsilon
\label{eqn:convergence_cyclic}
\ee
or $n \geq N_{co,1}$, where $\epsilon>0$ and $N_{co,1}$ is the maximum allowable number of iterations.
The proposed iterative algorithm is summarized in Algorithm \ref{alg1} and 
the adaptive modification of the LRT is given by
\be
\Psi_2(\bX)=\frac{{\ds
f_{X,1}(\bX;\widehat{\bm}^{(n)},(\widehat{\bsigma}^2_1)^{(n)})
}}
{{\ds
f_{X,1}(\bX;\bzero,\widehat{\bsigma}^2_0)
}}
\test\eta.
\label{eqn:adaptiveLRT_original}
\ee
The above architecture is referred to in what follows as Gaussian Detector for Heterogeneous Environment (GD-HE).

\subsection{Invariant Data Domain}
In this subsection, the design is conducted by invoking the Principle of Invariance and the LRT is function 
of transformed data, namely
\be
\Lambda_1\left(\bZ;\bm,\bsigma^2\right) = 
\frac{\ds 
f_1\left(\bZ;\bm,\bsigma^2\right) }
{\ds 
f_0\left(\bZ\right) }
\test \eta,
\label{eqn:LRT}
\ee
where $f_1\left(\bZ;\bm,\bsigma^2\right)=\prod_{k=1}^{K} f_1\left(\bz_k;\bm,\sigma^2_k\right)$,
$f_0\left(\bZ\right)=\prod_{k=1}^{K} f_0\left(\bz_k\right)$. 
As stated at the beginning of this section, in order to estimate $\bm$ and $\bsigma^2$, we follow a 
cyclic procedure where, at each step, the EM-Algorithm is exploited (in place of the MLA) under the 
fictitious assumption that $\bz_k$, $k=1,\dots,K$, represent the observed data, while missing data are the norms 
of $\bx_k$, $k=1,\ldots,K$. Finally, we refer to $\bx_k$, $k=1,\ldots,K$, as complete data. 
The considered procedure relies on the following steps
\begin{enumerate}
\item assume that $\bsigma^2$ is known and estimate of $\bm$ using the EM-Algorithm;
\item replace $\bm$ with the estimate obtained at the previous step and estimate $\bsigma^2$ applying 
the EM-Algorithm for known $\bm$;
\item repeat the above steps until a stopping criterion is satisfied.
\end{enumerate}

\subsubsection{First step of the cyclic procedure}
let us assume that $\bsigma^2$ is known and estimate $\bm$. 
To this end, observe that the distribution of the complete data belongs to the exponential 
family \cite{LehmannBook} and, hence,
the EM-Algorithm simplifies. 
As a matter of fact, with focus on the complete data, 
by the Fisher-Neyman Factorization Theorem \cite{scharf1991statistical}, a sufficient statistic for $\bm$ is given by 
$\bt(\bX;\bsigma) = \sum_{k=1}^{K}\frac{\bx_k}{\sigma^2_k} = \sum_{k=1}^{K}\frac{b_k \bz_k}{\sigma^2_k}$,
where $b_k=\|\bx_k\|$.
Then, the expectation step of the EM-Algorithm 
consists in computing the conditional expectation of the sufficient statistic given the observed data, namely
\be
E[\bt(\bX;\bsigma)|\bZ;\bm,\bsigma^2]=\sum_{k=1}^{K}\frac{E[b_k|\bz_k;\bm,\sigma_k^2] \bz_k}{\sigma^2_k}.
\label{eqn:condExp}
\ee
In order to evaluate $E[b_k|\bz_k;\bm,\sigma^2]$, the conditional pdf of $b_k$ given $\bz_k$ is required. 
To this end, exploiting the definition of conditional pdf, we obtain
\be
f(b_k | \bz_k ;\bm,\sigma^2_k) = {f(b_k , \bz_k ;\bm,\sigma^2_k)}/{f(\bz_k;\bm,\sigma^2_k)}.
\ee
The numerator of the last equation can be obtained by considering the pdf of $\bx_k$ and performing the following transformation
$x_{k,1} = b_k\;\cos\left(\theta_k\right)$ and $x_{k,2} = b_k\;\sin\left(\theta_k\right)$,
where $\bx_k = \left[x_{k,1} \  x_{k,2}\right]^T$. 
The Jacobian of the transformation is $b_k$ and, hence, the transformed pdf is given by
\begin{align}
&f\left(b_k,\bz_k;\bm,\sigma^2_k \right) = \frac{b_k}{2\pi\sigma_k^2}
\exp\left\{\frac{\left(b_k\bz_k-\bm\right)^T \left(b_k\bz_k-\bm\right)}{-2\sigma_k^2}\right\} \nonumber
\\
&=\frac{1}{2\pi\sigma_k^2}\exp\left\{\frac{\left(b_k^2+\|\bm\|^2-2 b_k \bm^T\bz_k \right)}{-2\sigma_k^2}+ \log\left(b_k\right)\right\}.
\label{eqn:transformationVar2}
\end{align}
Finally, $f(b_k | \bz_k ;\bm,\sigma^2_k)$ can be recast as
\begin{multline}
f(b_k | \bz_k ;\bm,\sigma^2_k) 
\\
= \exp\left\{- \frac{b_k^2}{2\sigma^2_k} + \frac{b_k}{\sigma^2_k} \bz_k^T\bm + \log\left(b_k\right) 
- \xi\left(\bz_k^T\bm\right)  \right\},  
\label{eqn:transformationVar3}
\end{multline}
where
\be
\xi\left(\bz_k^T\bm\right) = \log\left[\sigma^2_k+\sigma_k \bz_k^T \bm \frac{\Phi\left(\bz_k^T \bm/\sigma_k\right)}
{\varphi\left(\bz_k^T \bm/\sigma_k\right)}\right].
\label{eqn:transformationVar4}
\ee
Note that the distribution of the random variable $b_k|\bz_k$ belongs to the exponential family with natural scalar parameter $p_k=\bz_k^T\bm$ \cite{LehmannBook}
since the pdf \eqref{eqn:transformationVar3} can be rewritten as \cite{shao2008mathematical}
$f\left(b_k|\bz_k;\bm,\sigma^2_k\right) = \exp\left\{ t(b_k)p_k - \xi\left(p_k\right) \right\} h(b_k)$,  
where $t(b_k)=b_k/\sigma^2_k$, $\ h(b_k)= \exp\{\log\left(b_k\right)-{b_k^2}/{2\sigma^2_k}\}$, and (see Appendix \ref{App:proofXi} for the proof)
\be
\xi(p_k)=\log\left\{\int_{0}^{+\infty}\exp\{ t(b_k) p_k \}h(b_k)db_k\right\}.
\label{eqn:canonicalFormExp}
\ee
Following the lead of \cite{BrownExpFam} and \cite{MorrisExpFam}, it is possible to show that
\be
E[t(b_k)|\bz_k;\bm,\sigma_k^2] =  \left.\frac{d}{dp} \left[\xi\left(p\right)\right] \right|_{p = p_k}
\label{eqn:derivataExpFam}
\ee
and, hence, that
\begin{align}
&E[b_k|\bz_k;\bm,\sigma_k^2] = \left.\sigma^2_k\left\{\frac{d}{dp} \left[\xi\left(p\right)\right] \right\}\right|_{p = p_k}\nonumber
\\
&=\bz_k^T\bm+\frac{\sigma_k^2\Phi(\bz_k^T\bm/\sigma_k)}{\sigma_k\varphi(\bz_k^T\bm/\sigma_k)+\bz_i^T\bm\Phi(\bz_k^T\bm/\sigma_k)} \nonumber
\\
&=h_k(\bm).
\label{eqn:canonicalParam}
\end{align}
Let us denote by $\widehat{\bm}^{(n)}$ an estimate of $\bm$ at the $n$th EM iteration, 
then the maximization step of the EM-Algorithm specialized for the exponential family consists 
in solving the equations \cite[Section II]{Dempster77}
\begin{align}
&E[\bt(\bX;\bsigma);\bm,\bsigma^2] =  \sum_{k=1}^{K}\frac{h_k(\widehat{\bm}^{(n)}) \bz_k}{\sigma^2_k} \nonumber
\\
& \Rightarrow \widehat{\bm}^{(n+1)} = \left[ \sum_{k=1}^K \frac{1}{\sigma^2_k}  \right]^{-1}
\sum_{k=1}^{K}\frac{h_k(\widehat{\bm}^{(n)}) \bz_k}{\sigma^2_k}.
\label{eqn:maximizationStep01}
\end{align}
Summarizing, the EM proposed algorithm starts from an initial estimate $\widehat{\bm}^{(0)}$ and, 
at each iteration, updates the estimate according to 
equation \eqref{eqn:maximizationStep01}. The iterations terminate when
\be
\left| f_1(\bZ;\widehat{\bm}^{(n)},\bsigma) - f_1(\bZ;\widehat{\bm}^{(n-1)},\bsigma) \right| < \epsilon_1
\label{eqn:convergence_EM1}
\ee
or $n \geq N_{EM,m}$, where $\epsilon_1>0$ and $N_{EM,m}$ is the maximum allowable number of iterations 
for the EM-Algorithm.

\subsubsection{Second step of the cyclic procedure}
this step provides an estimate of $\bsigma^2$ assuming 
that $\bm$ is known (for instance, it can be equal to $\widehat{\bm}^{(n)}$) 
and $\sigma_k^2\geq C_0$, $k=1,\ldots,K$. 
In this case, a sufficient statistic for $\bsigma^2$ is given by
\begin{align}
&\bt(\bX) = [\|\bx_1-\bm\|^2,\ldots,\|\bx_K-\bm\|^2]^T \nonumber
\\
& = [\|b_1\bz_1-\bm\|2,\ldots,\|b_K\bz_K-\bm\|^2]^T
\end{align}
and its conditional expectation given $\bZ$ can be written as
\be
E[\bt(\bX)|\bZ;\bm,\bsigma^2]\!\!\!=\!\!\!
\begin{bmatrix}
E[\|\bx_1-\bm\|^2|\bz_1;\bm,\sigma_1^2]
\\
\vdots
\\
E[\|\bx_K-\bm\|^2|\bz_K;\bm,\sigma_K^2]
\end{bmatrix}.
\ee
Let us focus on the $k$th entry of the above vector and 
exploit \eqref{eqn:derivataExpFam} and \eqref{eqn:canonicalParam} to obtain that
\begin{align}
& E[\|\bx_k-\bm\|^2|\bz_k;\bm,\sigma^2_k] \nonumber
\\
&=E[b_k^2|\bz_k;\bm,\sigma_k^2]+\|\bm\|^2-2p_k \left.\sigma^2_k\left\{\frac{d}{dp} \left[\xi\left(p\right)\right] \right\}\right|_{p = p_k} \nonumber
\\
&=\Var[b_k|\bz_k;\bm,\sigma_k^2]+
\left\{\left.\sigma^2_k\left\{\frac{d}{dp} \left[\xi\left(p\right)\right] \right\}\right|_{p = p_k}\right\}^2
+\|\bm\|^2 \nonumber
\\
& -2p_k \left.\sigma^2_k\left\{\frac{d}{dp} \left[\xi\left(p\right)\right] \right\}\right|_{p = p_k}.
\label{eqn:E_stepSigma}
\end{align}
Again, from the properties of the exponential family \cite{BrownExpFam,MorrisExpFam}, 
it turns out that
\begin{align}
&\Var[t(b_k)|\bz_k;\bm,\sigma_k^2] = \frac{1}{(\sigma^2_k)^2}\Var[b_k|\bz_k;\bm,\sigma_k^2] \nonumber
\\
&= \left.\frac{d^2}{dp^2} \left[\xi\left(p\right)\right] \right|_{p = p_k} = \frac{1}{\sigma_k^2} 
+ \frac{\varphi(p_k/\sigma_k)/\sigma_k}{\sigma_k\varphi(p_k/\sigma_k)+p\Phi(p_k/\sigma_k)}\nonumber
\\
& -\left[\frac{\Phi(p_k/\sigma_k)}{\sigma_k\varphi(p_k/\sigma_k)+p\Phi(p_k/\sigma_k)}\right]^2.
\end{align}
Thus, replacing the above equation into \eqref{eqn:E_stepSigma}, we obtain that
$E[\|\bx_k-\bm\|^2|\bz_k;\bm,\sigma_k^2]=\frac{2\sigma_k^3 \varphi(p_k/\sigma_k)}{ \sigma_k \varphi(p_k/\sigma_k)+p_k \Phi(p_k/\sigma_k)) }
- [p_k]^2 + p_k \frac{\sigma_k^2 \varphi(p_k/\sigma_k) }{ \sigma_k \varphi(p_k/\sigma_k)+p_k \Phi(p_k/\sigma_k))} + \|\bm\|^2$.

Now, let us denote by $(\widehat{\sigma}^2_k)^{(n)}$ an estimate of $\sigma^2_k$ at the $n$th iteration of the EM-Algorithm and solve 
with respect to $\sigma^2_k$ the following equation
$E[\|\bx_k-\bm\|^2;\bm,\sigma_k^2]=
\frac{2(\sigma_k^{(n)})^3 \varphi(p_k/\sigma_k^{(n)})}{ \sigma_k \varphi(p_k/\sigma_k^{(n)})+p_k \Phi(p_k/\sigma_k^{(n)})) }- [p_k]^2 
+ p_k \frac{(\sigma_k^{(n)})^2 \varphi(p_k/\sigma_k^{(n)}) }{ \sigma_k^{(n)} \varphi(p_k/\sigma_k^{(n)})+p_k \Phi(p_k/\sigma_k^{(n)}))} 
+ \|\bm\|^2$
to come up with
\begin{multline}
(\tilde{\sigma}^2_k)^{(n+1)}=
\frac{(\sigma_k^{(n)})^3 \varphi(p_k/\sigma_k^{(n)})}{ \sigma_k \varphi(p_k/\sigma_k^{(n)})+p_k \Phi(p_k/\sigma_k^{(n)}) }
- \frac{[p_k]^2 }{2}
\\
+ \frac{p_k}{2} \frac{(\sigma_k^{(n)})^2 \varphi(p_k/\sigma_k^{(n)}) }{ \sigma_k^{(n)} \varphi(p_k/\sigma_k^{(n)})+p_k 
\Phi(p_k/\sigma_k^{(n)})} 
+ \frac{\|\bm\|^2}{2}.
\label{eqn:sigmaEstimateIterative}
\end{multline}
In order to fulfill the constraint on $\sigma_k^2$, which is required to avoid numerical instability,
we regularize the estimate of $\bsigma^2$ 
as $(\widehat{\sigma}^2_k)^{(n+1)}=\max\{ (\tilde{\sigma}^2_k)^{(n+1)},C_0 \}$.
The stopping condition for this step is given by
\be
\left| f_1(\bZ;{\bm},(\widehat{\bsigma}^2)^{(n)}) - f_1(\bZ;{\bm},(\widehat{\bsigma}^2)^{(n-1)}) \right| < \epsilon_2
\label{eqn:convergence_EM2}
\ee
or $n\geq N_{EM,\sigma}$ with $\epsilon_2>0$ and $N_{EM,m}$ the maximum allowable number of iterations, 
then the EM-Algorithm terminates.

Now, once $(\widehat{\bsigma}^2)^{(n)}$ is available, we can repeat the first 
step of the cyclic procedure exploiting the above estimate as initial value for $\bsigma^2$.
Note that this estimation procedure is ``doubly'' iterative, namely 
for each step of the cyclic procedure the EM algorithm is executed. 
For this reason, the estimates of $\bm$
and $\bsigma^2$ are denoted using a double superscript as $\widehat{\bm}^{(n),(i)}$ 
and $(\widehat{\bsigma}^2)^{(n),(i)}$, where $n$ indexes the EM iterations and $i$ refers to the iterations
of the cyclic procedure\footnote{Notice that in the derivations, the second index has been omitted in order not to burden the notation.}. 
Finally, the entire procedure, summarized in Algorithm \ref{alg2}, terminates when
\begin{multline}
\left| f_1(\bZ;\widehat{\bm}^{(n_i),(i)},(\widehat{\bsigma}^2)^{(n_i),(i)}) \right.
\\
\left. - f_1(\bZ;\widehat{\bm}^{(n_{i-1}),(i-1)},(\widehat{\bsigma}^2)^{(n_{i-1}),(i-1)}) \right| < \epsilon_3
\label{eqn:convergence_cyclicEM}
\end{multline}
or $i\geq N_{co,2}$, where $\epsilon_3>0$, $n_i$ is the number of EM 
iterations at the $i$th iteration of the cyclic procedure, 
and $N_{co,2}$ is the maximum allowable number of iterations.

\subsubsection{Likelihood Ratio Test}
Finally, replacing $\bm$ and $\bsigma^2$ in \eqref{eqn:LRT} with 
the respective estimates provided by Algorithm \ref{alg2} and taking the logarithm, we obtain the following decision rule
\begin{multline}
\Lambda_2(\bZ)=
-\|\widehat{\bm}\|^2\sum_{k=1}^K\frac{1}{2\widehat{\sigma}^2_k}
\\
+\sum_{k=1}^K\log\left[1+\frac{\ds\frac{\bz_k^T \widehat{\bm}}{\widehat{\sigma}_k}\;\; \Phi\left(\frac{\bz_k^T \widehat{\bm}}{\widehat{\sigma}_k}\right)}{\ds\varphi\left(\frac{\bz_k^T \widehat{\bm}}{\widehat{\sigma}_k}\right)}\right]\test\eta,
\label{eqn:testDetector}
\end{multline}
which will be referred to in the following as Angular-Gaussian Detector (AGD).

\subsection{Cross Architectures}
As stated at the beginning of this section, two additional architectures can be obtained by combining
\eqref{eqn:LRT_original} with the estimates provided by Algorithms \ref{alg2} and 
\eqref{eqn:LRT} with the estimates provided by Algorithm \ref{alg1}. These architectures
are referred to in the following as Cross GD-HE (C-GD-HE) and Cross AGD (C-AGD), respectively.

\medskip

Finally, we conclude this section by observing that
the conceived estimation procedures converge at least to a local stationary point. As a matter of fact,
it is clear that at each iteration of Algorithm \ref{alg1}, the likelihood increases \cite{Stoica_alternating}. 
As for the EM-based cyclic procedure, since at each iteration the EM returns (at least) a local stationary point, 
it is possible to obtain an increasing sequence of likelihood values at each iteration of 
the cyclic procedure that uses the EM algorithm. 


\section{Illustrative Examples and Discussion} 
\label{Sec:IllustrativeExamples}
In this section, we investigate the behaviors of the proposed decision schemes in terms of CFARness and 
detection performance. Specifically, this study is conducted using simulated data to assess the nominal behavior 
as well as real recorded data to evaluate the effectiveness of the proposed architectures when the operating scenario 
does not exactly match the design assumptions. 
Moreover, as preliminary step, a convergence analysis of the
estimation procedures is provided in order to justify the parameter choices.

In the next numerical examples, the iterative estimation procedure in the original domain starts by setting
$(\widehat{\bsigma}^2_1)^{(0)}=\|\bx_i\|^2$, whereas the EM-based procedure in 
the invariant domain begins from 
$\widehat{\bm}^{(0),(0)}=\frac{1}{K}\sum_{i=1}^K \bx_i$ and $(\tilde{\bsigma}^2)^{(0),(0)}=\frac{1}{2} 
\left\|\bx_i-\widehat{\bm}^{(0),(0)}\right\|^2$.

\subsection{Simulated Data} 
\label{Sec:SimulatedData}
The analysis presented in this subsection first determines the numbers of iterations for the estimation
procedures required to obtain satisfactory results. Then, it investigates
to what extent the $P_{fa}$ is sensitive 
to variations of the interference parameters when the thresholds are evaluated 
simulating white noise and for a nominal value of the $P_{fa}$. 
Finally, given a preassigned $P_{fa}$, the detection performance for different parameter settings 
are studied. 
All the numerical examples in this subsection are obtained by means of Monte Carlo counting techniques
based upon $100/P_{fa}$ and $10000$ independent trials to estimate the thresholds (or the $P_{fa}$) and 
the $P_d$, respectively.

The interference power is defined as
\be
\sigma_k^2=\Delta u_k + \sigma^2_n, \quad k=1,\ldots,K,
\label{eqn:interferenceModel}
\ee
where $\sigma^2_n=1$ is the noise power, $u_k\sim U(0,1)$, $k=1,\ldots,K$, and $\Delta$ represents the heterogeneity level, namely the greater its value, the more heterogeneous the interference. It is important to underline that a noninformative prior is exploited for the interference power.
Finally, all the illustrative examples assume $P_{fa}=10^{-2}$.

\begin{figure}[htp!]
\begin{center}
\includegraphics[width=7cm, height=4.5cm]{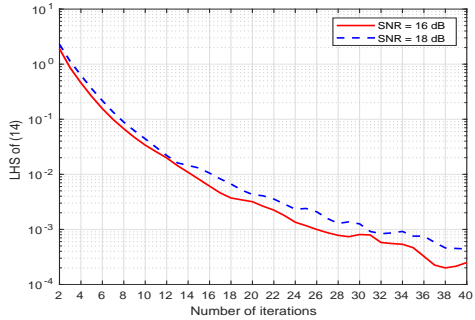}
\caption{LHS of \eqref{eqn:convergence_cyclic} versus the number of iterations (Algorithm 1).}
\label{fig:convergenceOriginal}
\end{center}
\end{figure}

\begin{figure}[htp!]
\begin{center}
\includegraphics[width=9cm, height=10cm]{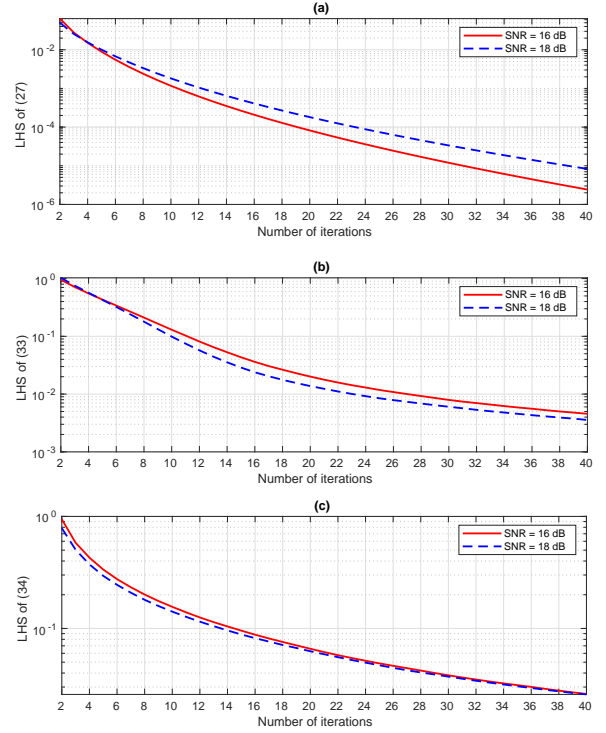}
\caption{Convergence curves for Algorithm 2: LHS of \eqref{eqn:convergence_EM1} versus the number of iterations (a); LHS of \eqref{eqn:convergence_EM2} versus the number of iterations (b); LHS of \eqref{eqn:convergence_cyclicEM} versus the number of iterations (c).}
\label{fig:convergenceInvariant}
\end{center}
\end{figure}

In order to select the number of iterations for the cyclic procedures and EM algorithm, in Figure \ref{fig:convergenceOriginal}-\ref{fig:convergenceInvariant}, we plot
the left-hand side (LHS) of \eqref{eqn:convergence_cyclic}, \eqref{eqn:convergence_EM1}, 
\eqref{eqn:convergence_EM2}, and \eqref{eqn:convergence_cyclicEM} versus the number of iterations 
for $K=16$ and $\Delta=10$. The two curves reported in the figures are related to two different SNR values
and are obtained by averaging over $10^4$ Monte Carlo trials. 
In Figure \ref{fig:convergenceOriginal}, we show the behavior of the stopping criterion given 
by \eqref{eqn:convergence_cyclic} for Algorithm 1. Inspection of the figure highlights that
a number of iterations greater than $15$ returns variations lower than $10^{-2}$. The next figure concerns
the convergence of Algorithm 2. Specifically, Subfigure \ref{fig:convergenceInvariant}(a) considers the 
LHS of \eqref{eqn:convergence_EM1}
where $\bsigma$ is replaced with the aforementioned initial value. It can be observed that $20$ iterations
are enough to appreciate a variation of the compressed likelihood less than $10^{-3}$. Now, we use this number
of iterations to obtain an initial estimate of $\bm$, which is, then, used to analyze the LHS of 
\eqref{eqn:convergence_EM2}. The resulting curve is plotted in Subfigure \ref{fig:convergenceInvariant}(b), 
where $20$ iterations provide a variation of about $10^{-2}$.
Finally, the curves reported in the third subfigure refer to the cyclic procedure of Algorithm 2 with 
$N_{EM,m}=N_{EM,\sigma}=20$. The subfigure points out that a number of iterations
greater than or equal to $15$ can represent a good compromise between computational complexity and
convergence issues. In a nutshell, in the next illustrative examples, we assume
$N_{co,1}=N_{co,2}=15$ and $N_{EM,m}=N_{EM,\sigma}=20$.

\begin{figure}[htp!]
\begin{center}
\includegraphics[scale=0.5]{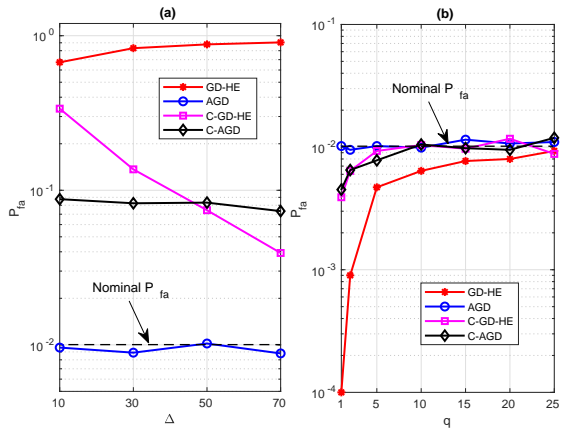}
\caption{(a) Estimated $P_{fa}$ versus $\Delta$ assuming model \eqref{eqn:interferenceModel}; (b)
Estimated $P_{fa}$ versus $q$; $K=16$ and thresholds computed under white noise hypothesis to ensure $P_{fa}=10^{-2}$.}
\label{fig:CFARanalysis}
\end{center}
\end{figure}

In Figure \ref{fig:CFARanalysis}(a), we estimate the $P_{fa}$ for the proposed 
detectors as a function of $\Delta$ (heterogeneity level) 
when the detection thresholds are 
computed assuming homogeneous white noise with power $\sigma^2_n$, $K=16$, and a nominal $P_{fa}=10^{-2}$. 
As expected, the AGD ensures the CFAR property since the estimated $P_{fa}$ is insensitive 
to the variations of $\Delta$ and is almost completely overlapped on the nominal $P_{fa}$.
As for the remaining detectors, the GD-HE exhibits a resulting $P_{fa}$ which is almost two orders of magnitude
higher than the nominal value ($10^{-2}$), whereas the $P_{fa}$ of C-AGD is close to $10^{-1}$. Finally,
the $P_{fa}$ curve related to C-GD-HE experiences a decreasing behavior.
In Figure \ref{fig:CFARanalysis}(b), we estimate the $P_{fa}$ assuming
a specific distribution for the interference power, i.e., data are modeled as
compound-Gaussian random variables \cite{conteCirculant,gini1}, namely
$x_k = \sqrt{\tau_k} g_k$, $k=1,\ldots,K$, where $g_k\sim \cC\cN_1(0,2\sigma^2_n)$ and $\tau_k$, $k=1,\ldots,K$, 
follows the Gamma distribution whose pdf
is $f(\tau_k)=\frac{\ds\tau_k^{b-1}}{\ds\beta^q\Gamma(q)}\exp\{ -\tau_k/\beta \}$
with $q>0$ and $\beta$ being the shape and scale parameters, respectively. The considered setting assumes $q=1/\beta$ to have a Gamma distribution with
unit mean. Observe that for large values of $q$, data distribution approaches the Gaussian distribution. 
The Figure highlights that
the $P_{fa}$ of the GD-HE, C-GD-HE, and C-AGD depends on the shape parameter $q$. 
Specifically, for low values of $q$, the estimated $P_{fa}$ significantly deviates from the nominal value 
On the other hand, as $q$ increases,
the environment tends to be homogeneous and, hence, the $P_{fa}$ of GD-HE, C-GD-HE, and C-AGD approaches 
the nominal value $10^{-2}$. 
As for the AGD, the estimated $P_{fa}$ is very close to the nominal $P_{fa}$ regardless of 
the shape parameter value. 

Summarizing, this analysis has corroborated that the AGD can ensure the CFAR property with respect to 
the power level of the interference in heterogeneous environments. On the contrary, the GD-HE, C-GD-HE, and C-AGD 
are not capable of maintain the false alarm rate constant. The last behavior can be explained by observing that 
the use of original data for estimation and/or detection does not allow to get rid of the dependence on the 
interference power at least for the aforementioned decision schemes, whereas the AGD takes advantage of the 
Invariance Principle to preserve the $P_{fa}$. 
Therefore, the GD-HE, C-GD-HE, and C-AGD do not
exhibit usage flexibility; they could be possibly exploited in conjunction with a clutter map 
and a lookup table for the detection threshold selection.

Now, we focus on the detection performance of the devised architectures assuming model \eqref{eqn:interferenceModel} and $P_{fa}=10^{-2}$;
the Signal-to-Noise Ratio (SNR) is defined as $\mbox{SNR}=\|\bm\|^2/\sigma^2_n$.
For comparison purposes, we also report the $P_d$ curves of the Clairvoyant Detector (CD) based 
upon the LRT, whose expression is\footnote{Note that this decision scheme cannot be used in practice since it assumes 
the perfect knowledge of $\bm$ and $\sigma_k$.}
\be
-\sum_{k=1}^{K}\frac{\| \bx_k-\bm \|^2}{\sigma^2_k}+\sum_{k=1}^{K}\frac{\| \bx_k \|^2}{\sigma^2_k}\test \eta,
\ee
the noncoherent linear detector or Energy Detector (ED) and the coherent detector (CHD),
whose expressions are
\be
\sum_{k=1}^{K}\| \bx_k \|^2\test \eta \quad \mbox{and} \quad \left\|\sum_{k=1}^{K} \bx_k \right\|^2\test \eta,
\ee
respectively.

\begin{figure}[htp!]
\begin{center}
\includegraphics[width=9cm, height=9.5cm]{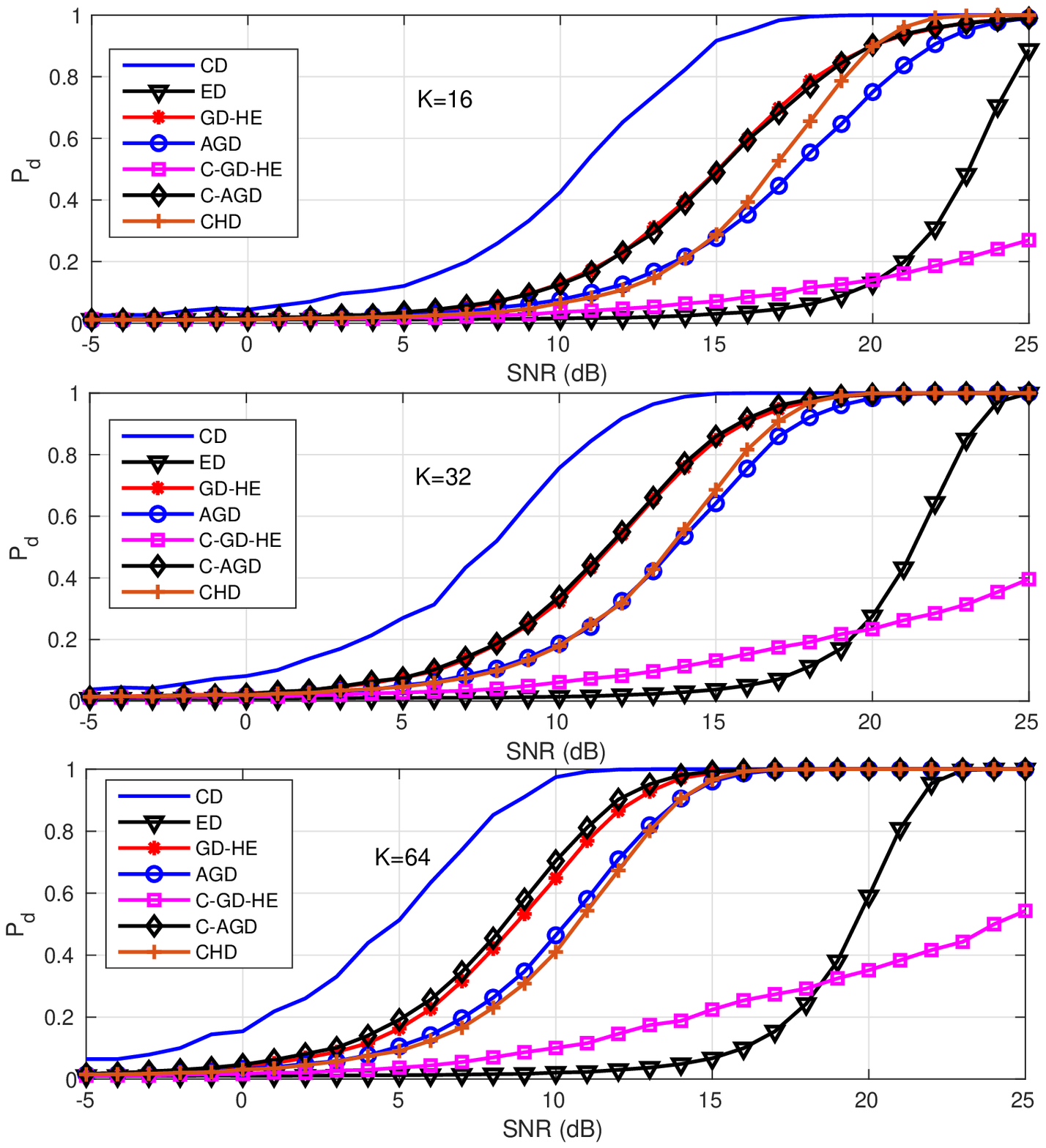}
\caption{$P_{d}$ versus SNR for the CD, ED, GD-HE, AGD, C-GD-HE, C-AGD, and CHD assuming $\Delta=10$ 
and $P_{fa}=10^{-2}$.}
\label{fig:Pd_Delta10}
\end{center}
\end{figure}

\begin{figure}[htp!]
\begin{center}
\includegraphics[width=8cm, height=3cm]{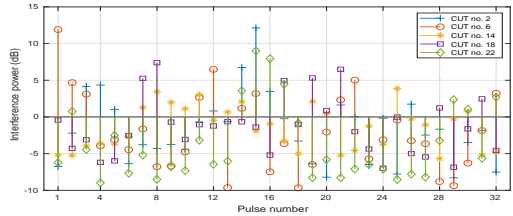}
\caption{Power variation over the pulse burst for some range bins.}
\label{fig:powerVariation}
\end{center}
\end{figure}

In Figure \ref{fig:Pd_Delta10}, we plot the $P_d$ curves for $\Delta=10$ 
and different values of $K$.
The value of $\Delta$ corresponds to a moderate heterogeneity level and 
leads to a CNR of about $9$ dB.
From inspection of the subfigures
it turns out that the GD-HE and C-AGD exhibit better detection performances than the AGD, ED, CHD, and C-GD-HE.
The latter is not capable to achieve $P_d=1$ for the considered parameter setting and its $P_d$ curves intersect
those of ED. The loss of the ED with respect to the AGD at $P_d=0.9$ 
increases from about $4.5$ dB for $K=16$ to about $8$ dB when $K=64$. 
The curves of the CHD are in between those of the GD-HE and of the AGD with a gain over the latter
that decreases as $K$ increases (note that for $K=32,64$ the considered curves are very close to each other).
Moreover, the 
GD-HE and C-AGD experience a gain in between $1.5$ dB (for $K=16$) and $2$ dB (for $K=64$) 
over the AGD (at $P_d=0.9$).
This hierarchy can be explained by the fact that the AGD is built up over normalized 
data and, hence, does not exploit all the available information with
an avoidable performance degradation due to a lower estimation quality. However, such information loss 
allows to gain the CFAR property as shown in the previous figures. 
On the contrary, the GD-HE and C-AGD take advantage of all the available 
information but, as already highlighted, they do not guarantee the CFARness, which is of primary concern in radar.
In the next figure, we compare the performances of the considered detectors assuming the same parameters 
as the previous figures but for $\Delta=50$, which leads to a more severe level of heterogeneity with 
respect to the previous examples,
since now the CNR increases to about $23$ dB. Fluctuations of this order of magnitude
can be observed in Figure \ref{fig:powerVariation} where clutter power variations over 
the time for live-recorded data are shown.

\begin{figure}[htp!]
\begin{center}
\includegraphics[width=9cm, height=9.5cm]{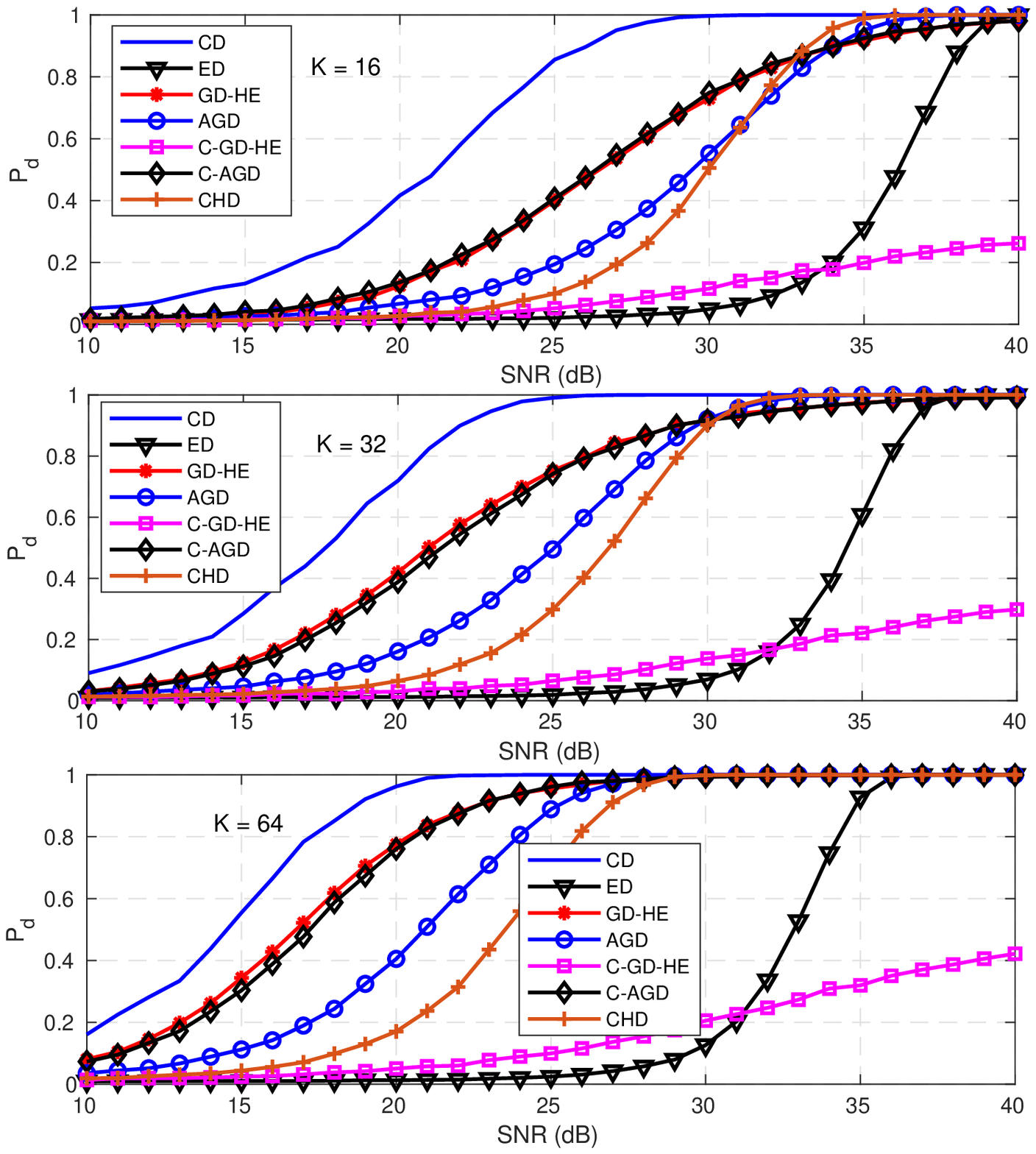}
\caption{$P_{d}$ versus SNR for the CD, ED, GD-HE, AGD, C-GD-HE, C-AGD, and CHD assuming $\Delta=50$ 
and $P_{fa}=10^{-2}$.}
\label{fig:Pd_Delta50}
\end{center}
\end{figure}

Figure \ref{fig:Pd_Delta50} confirms the behavior observed in Figure \ref{fig:Pd_Delta10} 
with the difference that there exists an intersection between the $P_d$ curves 
of the GD-HE (and C-AGD) with those of the AGD and CHD in the 
high SNR region, where the latter slightly outperform the former. Moreover, the curves of the AGD and CHD intersect
each other and the intersection point moves towards high SNR values as $K$ grows leading to a situation where 
the AGD outperforms the CHD with a gain of about $2$ dB at $P_d=0.9$ for $K=64$.
The ED detector provides poor performances with a loss at $P_d=0.9$ with respect to the AGD that increases to 
about $10$ dB for $K=64$, whereas, for the considered simulating scenario, the maximum $P_d$ value achieved by
the C-GD-HE is $0.4$ for $K=64$.

\begin{figure}[htp!]
\begin{center}
\includegraphics[width=9cm, height=9.5cm]{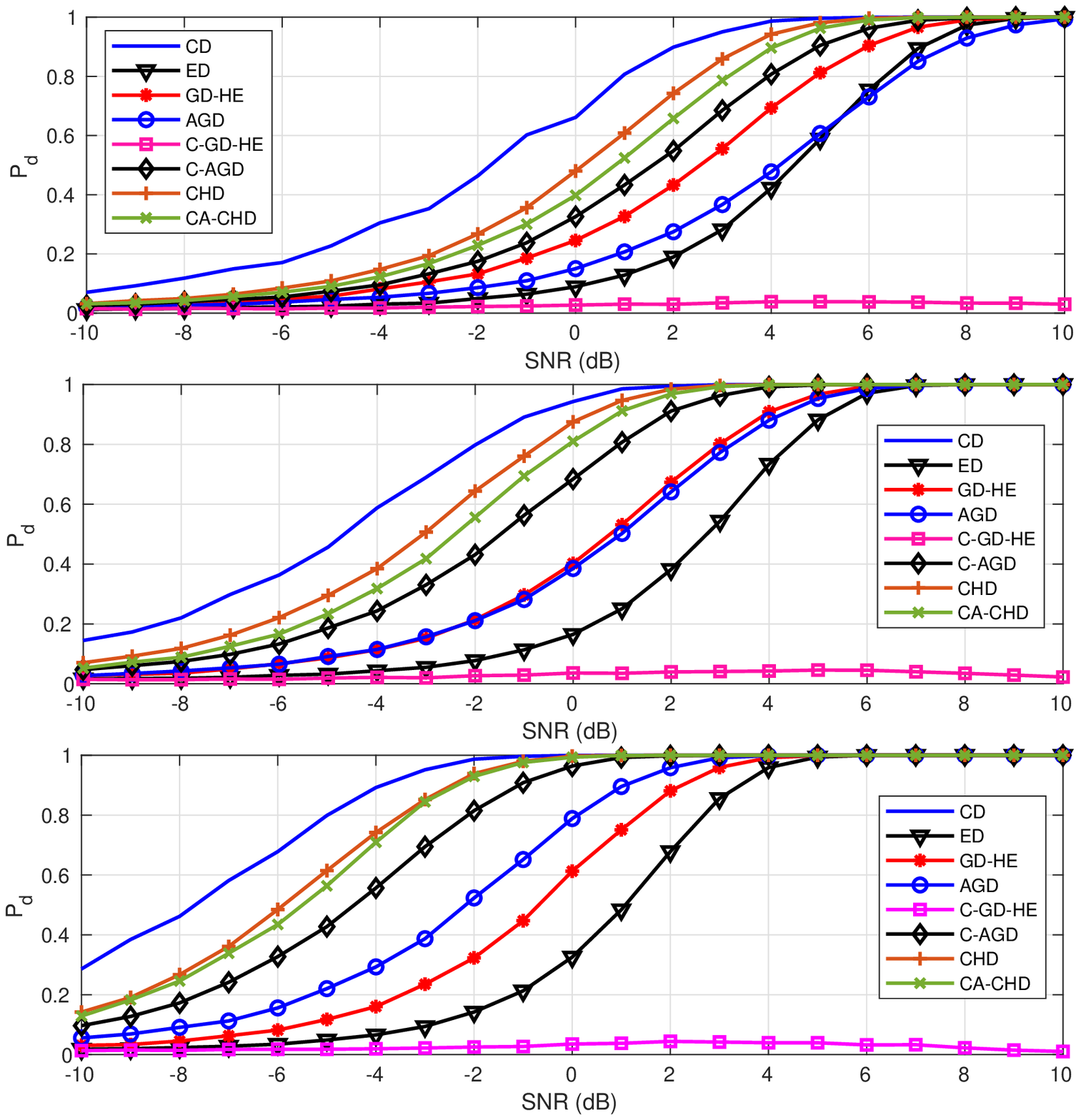}
\caption{$P_{d}$ versus SNR for the CD, ED, GD-HE, AGD, C-GD-HE, C-AGD, CHD, and CA-CHD assuming $\Delta=0$ 
(homogeneous environment) and $P_{fa}=10^{-2}$.}
\label{fig:Pd_Delta0}
\end{center}
\end{figure}

For completeness, in Figure \ref{fig:Pd_Delta0}, we show the performance of the new architectures when
$\Delta=0$, namely under the homogeneous environment. The curves of the cell-averaged coherent 
detector (CA-CHD), whose statistic is 
$\|\sum_{k=1}^{K} \bx_k \|^2 / \sum_{k=1}^{K}\| \bx_k \|^2$, are also reported. In this case, the CHD 
followed by the CA-CHD overcome the other detectors (except for the CD) with the CHD gaining about $0.5$ dB
with respect to the CA-CHD (the loss of the latter is due to the CFAR behavior \cite{Richards}). The AGD shares
almost the same performance as the ED for $K=16$, but as $K$ becomes larger and larger, the related $P_d$
curves improve. In fact, for $K=64$ and $P_d=0.9$, the AGD exhibits a loss of about $2$ dB with respect to the
C-AGD and a gain of more than $1$ dB over the GD-HE.

Thus, the analysis on simulated data has singled out the AGD as an effective means to deal with 
heterogeneous data, since it ensures reasonable detection performances and, at the same time, 
retains the CFAR property, which is of primary importance in radar.

\subsection{Real Data} 
\label{Sec:RealData}
In this section, we present numerical examples based upon live recorded data. To this end, we use the measurements 
which have been recorded in winter 1998 using the McMaster IPIX radar in Grimsby, on the shore of Lake Ontario, 
between Toronto
and Niagara Falls. Specifically, we test the proposed algorithm on dataset 85 for the 
HH polarization and in order to meet the requirement on the noise power lower bound, 
we add $1$ to data.
A detailed statistical analysis of the adopted real data has been conducted in \cite{conte2004statistical}.

\begin{figure}[htp!]
\begin{center}
\includegraphics[scale=0.5]{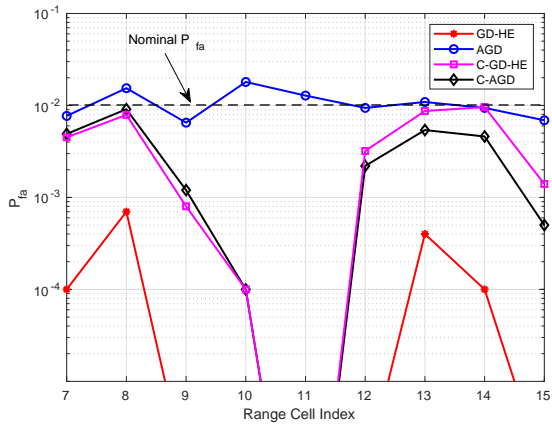}
\caption{$P_{fa}$ estimated from the $7$th to the $15$th range bin for AGD and GD-HE over IPIX data 
assuming $K=16$ and thresholds computed under white noise hypothesis to ensure $P_{fa}=10^{-2}$.}
\label{fig:CFARreal}
\end{center}
\end{figure}

The first analysis focuses on the CFARness and consists in estimating the $P_{fa}$ when the thresholds have 
been set under the white noise assumption with $C_0=1$.
Specifically, the nominal $P_{fa}$ is set to $10^{-2}$ and heterogeneous data are selected 
using a sliding mechanism to generate $100/P_{fa}$ sets of possibly uncorrelated samples 
(high pulse repetition intervals).
Before proceeding with the analysis, in Figure \ref{fig:powerVariation}, we provide a glimpse of the data nature 
in terms of power variations for some pulse bursts. The figure highlighted the presence of significant power 
variations over the pulses confirming the heterogeneous nature of data (other sets not considered here for brevity
experience an analogous behavior). The results of the CFAR analysis are shown in Figure \ref{fig:CFARreal}.
From the figure it turns out that the actual $P_{fa}$ of the AGD is very close to the nominal one confirming its CFAR behavior. 
On the other hand, the $P_{fa}$ of the remaining architectures considerably deviates 
from the nominal value. Specifically, the worst situation is experienced by the GD-HE, whose $P_{fa}$ values
are always below $10^{-3}$. As for the C-AGD and C-GD-HE, they exhibit $P_{fa}$ values close to the nominal at a 
few range indices. It is also important to notice that for these architectures, the discrepancy 
with respect to the nominal $P_{fa}$ values can achieve several orders of magnitude with values 
outside the range considered in Figure \ref{fig:CFARreal}.

\begin{figure}[htp!]
\begin{center}
\includegraphics[width=9cm, height=7.5cm]{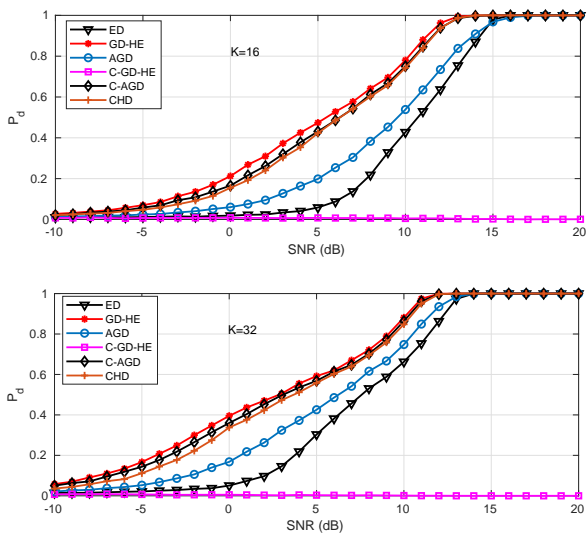}
\caption{$P_{d}$ versus SNR for the ED, GD-HE, AGD, C-GD-HE, C-AGD, and CHD over IPIX data assuming $P_{fa}=10^{-2}$.}
\label{fig:Pd_real}
\end{center}
\end{figure}

Finally, in Figure \ref{fig:Pd_real}, we show the detection performance for different values 
of $K$ and for the $8$th range bin. In this case, all the considered decision schemes exploit 
a detection threshold ensuring the same $P_{fa}=10^{-2}$ and evaluated over the real data. 
The data set at each trial is obtained through a sliding window as described for the CFAR analysis. 
The figure is somehow reminiscent of the situation observed for synthetic data, 
where the GD-HE and C-AGD share the same performance confirming their superiority over the AGD with
a gain that reduces to about $1$ dB for $K=32$. The main difference with respect to previous figures resides
in the fact that the curves of the CHD are very close to those of the GD-HE and C-AGD.
However, it is important to recall that such
architectures do not provide a CFAR behavior and, hence, setting their thresholds in practical scenarios
is not an easy task. Finally, the C-GD-HE continues to exhibit very poor performance at least for the
considered parameter setting.

\section{Conclusions} 
\label{Sec:Conclusions}
In this paper, four new detection architectures for heterogeneous Gaussian environments have been proposed 
and assessed. Specifically, the first detector relies on original data and uses the likelihood ratio as decision
statistic where the unknown target and interference parameters are estimated by means of a cyclic optimization
procedure. The second decision scheme transfers data into the invariant domain and exploits
normalized data, which are functionally invariant of scaling factors, to build up a CFAR decision scheme. 
Then, an alternating procedure incorporating the EM-Algorithm is devised to estimate the unknown 
parameters in the invariant domain. The remaining architectures have been obtained by combining
the estimation procedure for the original data domain with the detector for the invariant data domain and vice versa.
The behavior of these architectures has been first investigated resorting to simulated data adhering the design assumptions and, then, they have been tested on real recorded data. The analysis has singled out the second 
decision scheme based upon normalized data as the recommended solution for adaptive detection in heterogeneous
environments since it can guarantee the CFAR property and a limited detetion loss with respect to the other
architectures which exploit estimates based upon all the available information carried by data and whose
$P_{fa}$ is very sensitive to the interference power variations. 

Finally, it would be of interest investigating the behavior of the proposed architectures
in the presence of a mismatch for the noise power lower bound as well as extending the herein presented 
approach to the case of coherent processing through space-time data vectors 
sharing the same structure of the interference covariance matrix but different power levels.

\section*{Acknowledgments}
The authors are deeply in debt to Prof. S. Haykin 
and Dr. B. Currie, McMaster University, who have kindly provided the IPIX data. 
Moreover, the authors are very grateful to the anonymous Reviewers and Associate Editor for their
useful and constructive comments.
This work was in part supported by the National Key Research and Development Program of China (No. 2018YFB1801105), 
the National Natural Science Foundation of China under Grant (No. 61871469), 
the Youth Innovation Promotion Association CAS (CX2100060053), and 
the Fundamental Research Funds for the Central Universities under Grant WK2100000006.

\appendices

\section{Maximal Invariant Statistic for Scaling Transformations}
\label{maximal_inv_proof}
In this appendix, we prove that \eqref{eqn:maxinv} is a MIS with respect to $\cG$. To this end, we recall that $\bT(\cdot,\ldots,\cdot)$ is said to be a MIS 
if and only if
\be
\begin{cases}
\bT(\bx_1,\ldots,\bx_K)  =\bT(\bG(\bx_1,\ldots,\bx_K)),\quad\forall \bG\in\cG\,;
\\
\begin{aligned}
\bT(\bx_1,& \ldots,\bx_K) = \bT(\bar{\bx}_1,\ldots,\bar{\bx}_K)\Rightarrow\\
& \exists\, 
\bar{\bG}\in\cG\,:\,[\bx_1,\ldots,\bx_K]=\bar{\bG}[\bar{\bx}_1,\ldots,\bar{\bx}_K]\,.
\end{aligned}
\end{cases}
\label{eqn:MISdef}
\ee
The first property is evident since $\bT\left(\bc(1)\bx_1,\ldots,\bc(K)\bx_K\right) = \bT\left(\bx_1,\ldots,\bx_K\right)$, 
$\forall \bc\in\R_+^{K\times 1}$.
To prove the maximality, assume that $\bT(\bx_1,\ldots,\bx_K) =\bT(\bar{\bx}_1,\ldots,\bar{\bx}_K)$ and let
$\bar{\bc}=\left[ \frac{\|\bx_1\|}{\|\bar{\bx}_1\|}, \ldots, \frac{\|\bx_K\|}{\|\bar{\bx}_K\|}\right]^T\in\R_+^{K\times 1}$.
It follows that we can define the action 
$\bar{\bG}(\bar{\bx}_1,\ldots,\bar{\bx}_K) = [\bar{\bc}(1)\bar{\bx}_1,\ldots,\bar{\bc}(K)\bar{\bx}_K]
= [\bx_1,\ldots,\bx_K]$.
Thus, we have found $\bar{\bG}\in\cG$ which meets the second requirement of \eqref{eqn:MISdef} and the proof is complete.

\section{Derivation of \eqref{eqn:pdf_H1}}
\label{App:derivation_pdf_H1}
Let us start the derivation by writing (3.5.48) of \cite{MardiaJupp} with $\bSigma = \sigma^2_i\bI$, namely\footnote{Recall that $\|\bz_i\|=1$.}
\begin{align}
f_1(\bz_i;\bm,\sigma_i) &=\frac{1}{2\pi}\exp\left\{ \frac{-\|\bm\|^2}{(2\sigma^2_i)}\right\}+\frac{\bz_i^T\bm}{\sigma_i} 
\Phi\left(\frac{\bz_i^T\bm}{\sigma_i}\right) \nonumber
\\
&\times \frac{1}{\sqrt{2\pi}}
\exp\left\{ -\frac{(m_1 z_{i,2}-m_2 z_{i,1})^2}{2\sigma^2_i} \right\},
\label{eqn:pdf_00}
\end{align}
where $\bm=[m_1 \ m_2]^T$ and $\bz_i=[z_{i,1} \ z_{i,2}]^T$. Now, observe that 
\begin{align}
&(m_1 z_{i,2}-m_2 z_{i,1} )^2 = m_1^2 z_{i,2}^2 + m_2^2 z_{i,1}^2 - 2 m_1 m_2  z_{i,1} z_{i,2} \nonumber
\\
&= m_1^2 z_{i,2}^2 + m_2^2 z_{i,1}^2 - 2 m_1 m_2  z_{i,1} z_{i,2} + (\bz_i^T\bm)^2 - (\bz_i^T\bm)^2 \nonumber
\\
&= m_1^2 z_{i,2}^2 + m_2^2 z_{i,1}^2 + m_1^2 z_{i,1}^2 + m_2^2 z_{i,2}^2  - (\bz_i^T\bm)^2 \nonumber
\\
&= \|\bm\|^2 - (\bz_i^T\bm)^2.
\end{align}
Replacing the above result in \eqref{eqn:pdf_00}, we obtain \eqref{eqn:pdf_H1} and the proof is concluded.

\section{Expression of $\xi(p)$}
\label{App:proofXi}

Focus on the logarithm argument of \eqref{eqn:canonicalFormExp} and rewrite it as
\begin{align}
&\int_{0}^{+\infty}\!\!\!\!\!\!\exp\{ t(x_k) p_k \}h(x_k)dx_k = \int_{0}^{+\infty}\!\!\!\!\!\!\exp\left\{ \frac{x_k}{\sigma^2_k} p_k -\frac{x_k^2}{2\sigma^2_k}\right\}x_k dx_k  \nonumber
\\
&= \exp\left\{ \frac{p_k^2}{2\sigma^2_k} \right\} \int_{0}^{+\infty}\exp\left\{-\frac{(x_k-p_k)^2}{2\sigma^2_k} \right\}x_k dx_k \nonumber
\end{align}
\begin{align}
&=-\exp\left\{ \frac{p_k^2}{2\sigma^2_k} \right\} \sigma^2_k 
\int_{0}^{+\infty} 
\frac{{(p_k-x_k)}/{\sigma^2_k}}
{\exp\left\{{(x_k-p_k)^2}/{2\sigma^2_k} \right\}} dx_k \nonumber
\\
&+p_k \frac{\ds\int_{0}^{+\infty}\exp\left\{-\frac{(x_k-p_k)^2}{2\sigma^2_k} \right\} dx_k}
{\exp\left\{\ds-{p_k^2}/{2\sigma^2_k}\right\}} \nonumber
=\sigma^2_k+\sigma_k p_k \frac{ \Phi(p_k/\sigma_k)}{\varphi(p_k/\sigma_k)},
\end{align}
where the last equality concludes the proof.

\bibliographystyle{IEEEtran}
\bibliography{group_bib}

\begin{thebibliography}{10}
\providecommand{\url}[1]{#1}
\csname url@samestyle\endcsname
\providecommand{\newblock}{\relax}
\providecommand{\bibinfo}[2]{#2}
\providecommand{\BIBentrySTDinterwordspacing}{\spaceskip=0pt\relax}
\providecommand{\BIBentryALTinterwordstretchfactor}{4}
\providecommand{\BIBentryALTinterwordspacing}{\spaceskip=\fontdimen2\font plus
\BIBentryALTinterwordstretchfactor\fontdimen3\font minus
  \fontdimen4\font\relax}
\providecommand{\BIBforeignlanguage}[2]{{%
\expandafter\ifx\csname l@#1\endcsname\relax
\typeout{** WARNING: IEEEtran.bst: No hyphenation pattern has been}%
\typeout{** loaded for the language `#1'. Using the pattern for}%
\typeout{** the default language instead.}%
\else
\language=\csname l@#1\endcsname
\fi
#2}}
\providecommand{\BIBdecl}{\relax}
\BIBdecl

\bibitem{kelly1986adaptive}
E.~J. Kelly, ``An adaptive detection algorithm,'' \emph{IEEE Transactions on
  Aerospace and Electronic Systems}, no.~2, pp. 115--127, 1986.

\bibitem{robey1992cfar}
F.~C. Robey, D.~R. Fuhrmann, E.~J. Kelly, and R.~Nitzberg, ``{A {CFAR} adaptive
  matched filter detector},'' \emph{IEEE Transactions on Aerospace and
  Electronic Systems}, vol.~28, no.~1, pp. 208--216, 1992.

\bibitem{BOR-Morgan}
F.~Bandiera, D.~Orlando, and G.~Ricci, \emph{Advanced Radar Detection Schemes
  Under Mismatched Signal Models}, M.~. C.~P. Synthesis Lectures~on Signal
  Processing No.~8, Ed., San Rafael, US, 2009.

\bibitem{JunLiu00}
J.~Liu, W.~Liu, B.~Chen, H.~Liu, H.~Li, and C.~Hao, ``{Modified Rao Test for
  Multichannel Adaptive Signal Detection},'' \emph{IEEE Transactions on Signal
  Processing}, vol.~64, no.~3, pp. 714--725, 2016.

\bibitem{CP00}
C.~Hao, S.~Gazor, G.~Foglia, B.~Liu, and C.~Hou, ``Persymmetric adaptive
  detection and range estimation of a small target,'' \emph{IEEE Transactions
  on Aerospace and Electronic Systems}, vol.~51, no.~4, pp. 2590--2604, 2015.

\bibitem{HaoSP_HE}
C.~Hao, D.~Orlando, G.~Foglia, and G.~Giunta, ``Knowledge-based adaptive
  detection: Joint exploitation of clutter and system symmetry properties,''
  \emph{IEEE Signal Processing Letters}, vol.~23, no.~10, pp. 1489--1493,
  October 2016.

\bibitem{Richards}
M.~A. Richards, J.~A. Scheer, and W.~A. Holm, \emph{Principles of Modern Radar:
  Basic Principles}.\hskip 1em plus 0.5em minus 0.4em\relax Raleigh, NC:
  Scitech Publishing, 2010.

\bibitem{DD}
F.~Bandiera, O.~Besson, D.~Orlando, G.~Ricci, and L.~L. Scharf, ``{GLRT-Based
  Direction Detectors in Homogeneous Noise and Subspace Interference},''
  \emph{IEEE Transactions on Signal Processing}, vol.~55, no.~6, pp.
  2386--2394, June 2007.

\bibitem{Ciuonzo2}
D.~Ciuonzo, A.~De~Maio, and D.~Orlando, ``A unifying framework for adaptive
  radar detection in homogeneous plus structured interference-part ii:
  Detectors design,'' \emph{IEEE Transactions on Signal Processing}, vol.~64,
  no.~11, pp. 2907--2919, June 2016.

\bibitem{Raghavan}
R.~S. Raghavan, N.~Pulsone, and D.~J. McLaughlin, ``{Performance of the GLRT
  for adaptive vector subspace detection},'' \emph{IEEE Transactions on
  Aerospace and Electronic Systems}, vol.~32, no.~4, pp. 1473--1487, October
  1996.

\bibitem{reed1974rapid}
I.~S. Reed, J.~D. Mallett, and L.~E. Brennan, ``Rapid convergence rate in
  adaptive arrays,'' \emph{IEEE Transactions on Aerospace and Electronic
  Systems}, no.~6, pp. 853--863, 1974.

\bibitem{Melvin-2000}
W.~L. Melvin, ``Space-time adaptive radar performance in heterogeneous
  clutter,'' \emph{IEEE Transactions on Aerospace and Electronic Systems},
  vol.~36, no.~2, pp. 621--633, 2000.

\bibitem{bergin2002gmti}
J.~S. Bergin, P.~M. Techau, W.~L. Melvin, and J.~R. Guerci, ``{GMTI STAP} in
  target-rich environments: site-specific analysis,'' in \emph{Radar
  Conference, 2002. Proceedings of the IEEE}.\hskip 1em plus 0.5em minus
  0.4em\relax IEEE, 2002, pp. 391--396.

\bibitem{Hongbin}
P.~Wang, H.~Li, and B.~Himed, ``Knowledge-aided parametric tests for
  multichannel adaptive signal detection,'' \emph{IEEE Transactions on Signal
  Processing}, vol.~59, no.~12, pp. 5970--5982, 2011.

\bibitem{DeMaioFoglia01}
A.~De~Maio, A.~Farina, and G.~Foglia, ``Knowledge-aided bayesian radar
  detectors \& their application to live data,'' \emph{IEEE Transactions on
  Aerospace and Electronic Systems}, vol.~46, no.~1, pp. 170--183, 2010.

\bibitem{YuriJohnson}
Y.~I. {Abramovich} and B.~A. {Johnson}, ``Glrt-based detection-estimation for
  undersampled training conditions,'' \emph{IEEE Transactions on Signal
  Processing}, vol.~56, no.~8, pp. 3600--3612, Aug 2008.

\bibitem{yuriBesson}
Y.~I. {Abramovich} and O.~{Besson}, ``{On the Expected Likelihood Approach for
  Assessment of Regularization Covariance Matrix},'' \emph{IEEE Signal
  Processing Letters}, vol.~22, no.~6, pp. 777--781, June 2015.

\bibitem{WieselHero}
Y.~{Chen}, A.~{Wiesel}, and A.~O. {Hero}, ``Robust shrinkage estimation of
  high-dimensional covariance matrices,'' \emph{IEEE Transactions on Signal
  Processing}, vol.~59, no.~9, pp. 4097--4107, Sep. 2011.

\bibitem{Tyler}
E.~{Ollila} and D.~E. {Tyler}, ``Regularized$m$-estimators of scatter matrix,''
  \emph{IEEE Transactions on Signal Processing}, vol.~62, no.~22, pp.
  6059--6070, Nov 2014.

\bibitem{yuriModified}
Y.~I. {Abramovich}, N.~K. {Spencer}, and A.~Y. {Gorokhov}, ``Modified glrt and
  amf framework for adaptive detectors,'' \emph{IEEE Transactions on Aerospace
  and Electronic Systems}, vol.~43, no.~3, pp. 1017--1051, July 2007.

\bibitem{gerlach1}
M.~J. {Steiner} and K.~{Gerlach}, ``Fast converging adaptive processor or a
  structured covariance matrix,'' \emph{IEEE Transactions on Aerospace and
  Electronic Systems}, vol.~36, no.~4, pp. 1115--1126, Oct 2000.

\bibitem{629144}
M.~C. {Wicks}, W.~L. {Melvin}, and P.~{Chen}, ``An efficient architecture for
  nonhomogeneity detection in space-time adaptive processing airborne early
  warning radar,'' in \emph{Radar 97 (Conf. Publ. No. 449)}, Oct 1997, pp.
  295--299.

\bibitem{1191096}
M.~{Rangaswamy}, ``Non-homogeneity detector for gaussian and non-gaussian
  interference scenarios,'' in \emph{Sensor Array and Multichannel Signal
  Processing Workshop Proceedings, 2002}, Aug 2002, pp. 528--532.

\bibitem{767347}
R.~S. {Adve}, T.~B. {Hale}, and M.~C. {Wicks}, ``Transform domain localized
  processing using measured steering vectors and non-homogeneity detection,''
  in \emph{Proceedings of the 1999 IEEE Radar Conference. Radar into the Next
  Millennium (Cat. No.99CH36249)}, April 1999, pp. 285--290.

\bibitem{7575566}
L.~{Jiang} and T.~{Wang}, ``Robust non-homogeneity detector based on reweighted
  adaptive power residue,'' \emph{IET Radar, Sonar Navigation}, vol.~10, no.~8,
  pp. 1367--1374, 2016.

\bibitem{851934}
B.~{Himed}, Y.~{Salama}, and J.~H. {Michels}, ``Improved detection of close
  proximity targets using two-step nhd,'' in \emph{Record of the IEEE 2000
  International Radar Conference [Cat. No. 00CH37037]}, May 2000, pp. 781--786.

\bibitem{1433140}
M.~{Rangaswamy}, ``Statistical analysis of the nonhomogeneity detector for
  non-gaussian interference backgrounds,'' \emph{IEEE Transactions on Signal
  Processing}, vol.~53, no.~6, pp. 2101--2111, June 2005.

\bibitem{kraut1999cfar}
S.~Kraut and L.~L. Scharf, ``{The CFAR adaptive subspace detector is a
  scale-invariant GLRT},'' \emph{IEEE Transactions on Signal Processing},
  vol.~47, no.~9, pp. 2538--2541, 1999.

\bibitem{Ward1994}
J.~Ward, ``Space-time adaptive processing for airborne radar,'' MIT Lincoln
  Laboratory, Tech. Rep. 1015, 1994.

\bibitem{papoulis2002probability}
A.~Papoulis and S.~Pillai, \emph{{Probability, Random Variables, and Stochastic
  Processes}}.\hskip 1em plus 0.5em minus 0.4em\relax McGraw-Hill, 2002.

\bibitem{LehmannBook}
E.~L. Lehmann, \emph{Testing Statistical Hypotheses}, 2nd~ed.\hskip 1em plus
  0.5em minus 0.4em\relax New York, USA: Springer-Verlag, 1986.

\bibitem{MardiaJupp}
K.~V. Mardia and P.~E. Jupp, \emph{Directional Statistics}.\hskip 1em plus
  0.5em minus 0.4em\relax John Wiley \& Sons, 2000.

\bibitem{Dempster77}
A.~P. Dempster, N.~M. Laird, and D.~B. Rubin, ``Maximum likelihood from
  incomplete data via the {EM} algorithm,'' \emph{Journal of the Royal
  Statistical Society (Series B - Methodological)}, vol.~39, no.~1, pp. 1--38,
  1977.

\bibitem{haykin2002uncovering}
S.~Haykin, R.~Bakker, and B.~W. Currie, ``Uncovering nonlinear dynamics-the
  case study of sea clutter,'' \emph{Proceedings of the IEEE}, vol.~90, no.~5,
  pp. 860--881, 2002.

\bibitem{antennaBased}
A.~Farina, \emph{Antenna-Based Signal Processing Techniques for Radar Systems},
  A.~House, Ed., Boston, MA, 1992.

\bibitem{Stoica_alternating}
P.~Stoica and Y.~Selen, ``Cyclic minimizers, majorization techniques, and the
  expectation-maximization algorithm: a refresher,'' \emph{IEEE Signal
  Processing Magazine}, vol.~21, no.~1, pp. 112--114, 2004.

\bibitem{scharf1991statistical}
L.~Scharf and C.~Demeure, \emph{Statistical Signal Processing: Detection,
  Estimation, and Time Series Analysis}, ser. Addison-Wesley series in
  electrical and computer engineering.\hskip 1em plus 0.5em minus 0.4em\relax
  Addison-Wesley Publishing Company, 1991.

\bibitem{shao2008mathematical}
J.~Shao, \emph{Mathematical Statistics}, ser. Springer Texts in
  Statistics.\hskip 1em plus 0.5em minus 0.4em\relax Springer New York, 2008.

\bibitem{BrownExpFam}
L.~D. Brown, ``Fundamentals of statistical exponential families with
  applications in statistical decision theory,'' \emph{Lecture Notes-Monograph
  Series}, vol.~9, pp. i--279, 1986.

\bibitem{MorrisExpFam}
C.~N. Morris, ``Natural exponential families with quadratic variance functions:
  Statistical theory,'' \emph{The Annals of Statistics}, vol.~11, no.~2, pp.
  515--529, 1983.

\bibitem{conteCirculant}
E.~Conte, M.~Lops, and G.~Ricci, ``Adaptive detection schemes in
  {compound-Gaussian} clutter,'' \emph{IEEE Transactions on Aerospace and
  Electronic Systems}, vol.~34, no.~4, pp. 1058--1069, 1998.

\bibitem{gini1}
F.~Gini and A.~Farina, ``{Vector Subspace Detection in Compound-Gaussian
  Clutter Part I: Survey and New Results},'' \emph{IEEE Transactions on
  Aerospace and Electronic Systems}, vol.~38, no.~4, pp. 1295--1311, 2002.

\bibitem{conte2004statistical}
E.~Conte, A.~De~Maio, and C.~Galdi, ``Statistical analysis of real clutter at
  different range resolutions,'' \emph{IEEE Transactions on Aerospace and
  Electronic Systems}, vol.~40, no.~3, pp. 903--918, 2004.

\end{thebibliography}
\begin{algorithm}
\caption{Iterative estimation of $\bm$ and $\bsigma^2$ in the original domain}
\label{alg1}
\begin{algorithmic}[1]
    \REQUIRE $C_0$, $N_{co,1}$, $\epsilon$, $\bX$, and $(\bsigma^2)^{(0)}$.
    \ENSURE $\widehat{\bm}^{(n)}$ and $(\widehat{\bsigma}^2_1)^{(n)}$.
    \STATE Set $n=0$
    \STATE Set $n=n+1$
    \STATE Compute $\widehat{\bm}^{(n)}=\left[ \sum_{k=1}^K \frac{1}{(\sigma_k^2)^{(n-1)}} \right]^{-1} \sum_{k=1}^K\frac{{\bf x}_k}{(\sigma_k^2)^{(n-1)}}$
    \STATE Compute $(\widehat{\bsigma}^2_1)^{(n)}=[\max\{\frac{1}{2}\|\bx_1-\widehat{\bm}^{(n)}\|^2,C_0\},\ldots,
    \max\{\frac{1}{2}\|\bx_K-\widehat{\bm}^{(n)}\|^2,C_0\}]^T$
    \STATE If the stopping criterion is not satisfied go to step 2 else go to step 6
    \STATE Return $\widehat{\bm}^{(n)}$ and $(\widehat{\bsigma}^2_1)^{(n)}$
\end{algorithmic}
\end{algorithm}
\begin{algorithm}
\caption{Iterative estimation of $\bm$ and $\bsigma^2$ in the invariant domain}
\label{alg2}
\begin{algorithmic}[1]
    \REQUIRE $N_{EM,m}$, $N_{EM,\sigma}$, $N_{co,2}$, $C_0$, $\epsilon_1$, $\epsilon_2$, $\epsilon_3$,
    $\bz_k$, $k=1,\ldots,K$, $\widehat{\bm}^{(0),(0)}$, and $(\tilde{\bsigma}^2)^{(0),(0)}$.
    \ENSURE $\widehat{\bm}$ and $\tilde{\bsigma}$.
    \STATE Set $i=0$ and $\bar{\bsigma}^2=(\tilde{\bsigma}^2)^{(0),(0)}$
    \STATE Set $n=0$ 
    \STATE Compute $\forall k=1,\ldots,K$, $h_k^{(n)}=E[b_k|\bz_k;\widehat{\bm}^{(n),(i)},\bar{\bsigma}^2]$ 
    using \eqref{eqn:canonicalParam} and $\bH^{(n)} = \diag(h_1^{(n)},\ldots,h_K^{(n)})$
    \STATE Compute $\widehat{\bm}^{(n+1),(i)}$ using \eqref{eqn:maximizationStep01}
    \STATE Set $n=n+1$
    \STATE if the stopping criterion for the EM-Algorithm is satisfied go to step 7 else go to step 3
    \STATE Set $\bar{\bm}=\bm^{(n),(i)}$
    \STATE Set $n=0$ and $\bar{p}_k=\bz_k^T\bar{\bm}$ 
    \STATE Compute $(\tilde{\sigma}^2_k)^{(n+1),(i)}$ using \eqref{eqn:sigmaEstimateIterative} 
    and set  $(\widehat{\sigma}^2_k)^{(n+1),(i)}
    =\max\{ (\tilde{\sigma}^2_k)^{(n+1),(i)}, C_0 \}$
    \STATE Set $n=n+1$
    \STATE if the stopping criterion for the EM-Algorithm is satisfied set $i=i+1$ and go to step 12 else go to step 9
    \STATE if the stopping criterion for the cyclic procedure is not satisfied set $\bm^{(0),(i)}=\bar{\bm}$, 
    $\bar{\bsigma}^2=[(\widehat{\sigma}^2_1)^{(n),(i)},\ldots,(\widehat{\sigma}^2_K)^{(n),(i)}]^T$, and go to step 2,
    else return $\widehat{\bm}=\bar{\bm}$ 
    and $\widehat{\bsigma}^2=[(\widehat{\sigma}_1^2)^{(n),(i)},\ldots,(\widehat{\sigma}_K^2)^{(n),(i)}]^T$
\end{algorithmic}
\end{algorithm}
\end{document}